\documentclass[nolinenumbers]{aastex631}

	\usepackage{tabularx}
	\usepackage{multirow}

\begin{document}
			
			\title{A New Period Determination Method for Periodic Variable Stars}

			\author{Xiao-Hui Xu}
			\footnote{Corresponding author: Xiao-Hui Xu, Qing-Feng Zhu}
			\affiliation{Deep Space Exploration Laboratory / Department of Astronomy, University of Science and Technology of China, Hefei 230026, China}
			\affiliation{School of Astronomy and Space Sciences, University of Science and Technology of China, Hefei 230026, China}
			\affiliation{CAS Key Laboratory for Research in Galaxies and Cosmology, University of Science and Technology of China, Hefei 230026, China}
			
			\author{Qing-Feng Zhu}
			\footnote{E-mail: xiaohuixu@mail.ustc.edu.cn, zhuqf@ustc.edu.cn}
			\affiliation{Deep Space Exploration Laboratory / Department of Astronomy, University of Science and Technology of China, Hefei 230026, China}
			\affiliation{School of Astronomy and Space Sciences, University of Science and Technology of China, Hefei 230026, China}
			\affiliation{CAS Key Laboratory for Research in Galaxies and Cosmology, University of Science and Technology of China, Hefei 230026, China}
			
			\author{Xu-Zhi Li}
			\affiliation{Deep Space Exploration Laboratory / Department of Astronomy, University of Science and Technology of China, Hefei 230026, China}
			\affiliation{School of Astronomy and Space Sciences, University of Science and Technology of China, Hefei 230026, China}
			\affiliation{CAS Key Laboratory for Research in Galaxies and Cosmology, University of Science and Technology of China, Hefei 230026, China}
			
			\author{Bin Li}
			\affiliation{School of Astronomy and Space Sciences, University of Science and Technology of China, Hefei 230026, China}
			\affiliation{CAS Key Laboratory of Planetary Sciences, Purple Mountain Observatory, Chinese Academy of Sciences, Nanjing 210023, China}
			
			\author{Hang Zheng}
			\affiliation{Deep Space Exploration Laboratory / Department of Astronomy, University of Science and Technology of China, Hefei 230026, China}
			\affiliation{School of Astronomy and Space Sciences, University of Science and Technology of China, Hefei 230026, China}
			\affiliation{CAS Key Laboratory for Research in Galaxies and Cosmology, University of Science and Technology of China, Hefei 230026, China}
			
			\author{Jin-Sheng Qiu}
			\affiliation{Deep Space Exploration Laboratory / Department of Astronomy, University of Science and Technology of China, Hefei 230026, China}
			\affiliation{School of Astronomy and Space Sciences, University of Science and Technology of China, Hefei 230026, China}
			\affiliation{CAS Key Laboratory for Research in Galaxies and Cosmology, University of Science and Technology of China, Hefei 230026, China}
			
			\author{Hai-Bin Zhao}
			\affiliation{School of Astronomy and Space Sciences, University of Science and Technology of China, Hefei 230026, China}
			\affiliation{CAS Key Laboratory of Planetary Sciences, Purple Mountain Observatory, Chinese Academy of Sciences, Nanjing 210023, China}
			\affiliation{CAS Center for Excellence in Comparative Planetology, Chinese Academy of Sciences, Hefei 230026, China}
			

			
			
			\begin{abstract}
				
				Variable stars play a key role in understanding the Milky Way and the universe. The era of astronomical big data presents new challenges for quick identification of interesting and important variable stars. Accurately estimating the periods is the most important step to distinguish different types of variable stars. Here, we propose a new method of determining the variability periods. By combining the statistical parameters of the light curves, the colors of the variables, the window function and the Generalized Lomb–Scargle (GLS) algorithm, the aperiodic variables are excluded and the periodic variables are divided into eclipsing binaries and NEB variables (other types of periodic variable stars other than eclipsing binaries), the periods of the two main types of variables are derived. We construct a random forest classifier based on 241,154 periodic variables from the ASAS-SN and OGLE datasets of variables. The random forest classifier is trained on 17 features, among which 11 are extracted from the light curves and 6 are from the Gaia Early DR3, ALLWISE, and 2MASS catalogs. The variables are classified into 7 superclasses and 17 subclasses. In comparison with the ASAS-SN and OGLE catalogs, the classification accuracy is generally above approximately 82\% and the period accuracy is 70\%-99\%. To further test the reliability of the new method and classifier, we compare our results with the results of \cite{Chen2020}  for ZTF DR2. The classification accuracy is generally above 70\%. The period accuracy of the EW and SR variables is $\sim$ 50\% and 53\%, respectively. And the period accuracy of other types of variables is 65\%-98\%. 
				
			\end{abstract}
			
		
		
		\section{Introduction} \label{sec:packages1}
		
		Variable stars are a large and important branch of time-varying sources that help us understand the Milky Way and the universe. The Eclipsing binaries, especially detached systems with low mass components (with \emph{M$_{\star}$} $\leq$ 0.7M$_{\sun}$), are vital for testing stellar evolutionary models concerning the role of convection in late-type stars \citep{Feiden&Chaboyer2012,Han2019}. Cepheids have well-defined period-luminosity relations \citep{Breuval2020,Ripepi2020} and become primary distance indicators that can be used to estimate the Hubble constant \citep{Riess2011,Riess2019,PlanckCollaboration2020}. Over 4300 new RRab stars in \emph{K$_{s}$}-band toward the inner bulge were identified by \cite{Dekany&Grebel2020}, and most of them were beyond the detection limit of optical surveys, which extend the census of RRab stars to near the midplane of the Milky Way and greatly contribute to understanding the properties of interstellar extinction along the Galactic plane. \cite{Jayasinghe2020c} derived the period-luminosity relationship of Delta Scuti stars based on data from the All-Sky Automated Survey for Supernovae \citep[ASAS-SN;][]{Shappee2014,Jayasinghe2018a,Jayasinghe2018c} and found that stars with \emph{P} $>$ 0.1 d were mainly located near the Galactic disk ($|$Z$|$ $<$ 0.5 kpc). They further suggested that the radial period gradient, the median period increasing with the Galactocentric radius R at low latitudes, may imply a radial metallicity gradient. Miras can lead to the presence of elements heavier than nickel in the interstellar medium through mass loss phenomena \citep{Maciel1976,Wood&Cahn1977, Perrin2020}. Miras are also important distance indicators and tracers of galactic chemistry, stellar populations, and structure of galaxies \citep{Battistini&Bensby2016,Menzies2019,Huang2020,Yu2021}. 
		
		Ongoing or upcoming surveys, such as ASAS-SN, Optical Gravitational Lensing Experiment \citep[OGLE;][]{Udalski2003}, Zwicky Transient Facility \citep[ZTF;][]{Masci2019,Bellm2019}, China Near-Earth Object Survey Telescope \citep[CNEOST;][]{Wang2016,Yeh2020}, Gaia \citep{GaiaCollaboration2016,GaiaCollaboration2021,Mowlavi2018}, Legacy Survey of Space and Time \citep[LSST;][]{Ivezic2019}, and Wide Field Survey Telescope (WFST), will observe the light curves of billions of astronomical sources, and the large amount of data presents new challenges for the rapid identification and classification of different variability types \citep{Kim2021}.
		
		The OGLE project, which began in 1992 and was suspended in 2020 due to the COVID-19 pandemic, is one of the longest-lasting variable sky surveys in the world \citep{Udalski1992,Udalski1997,Udalski2015,Udalski2003}. The observations were carried out in the \emph{I}-band with the mean magnitudes from 10 mag to 20 mag and in the \emph{V}-band with the mean magnitudes from 13 mag to 23 mag \citep{Iwanek2022}. \cite{Iwanek2022} presented the epoch number statistics of the light curves of two bands in four phases of the OGLE project, and the maximum epochs of \emph{I}- and \emph{V}-band observations are 16,686 and 230, respectively. To date, the OGLE database\footnote{\url{https://ogle.astrouw.edu.pl}} has revealed approximately 730,000 variable stars.  
		
		The ASAS-SN is the first optical survey to monitor the entire sky, originally with a cadence of 2-3 days down to \emph{V} $\sim$ 17 mag, currently with a cadence of one day down to \emph{g} $\sim$ 18.5 mag, and has collected $\sim$ 2,000 to over 7,500 epochs of \emph{V}- and \emph{g}-band observations per field to date \citep{Jayasinghe2018c,Christy2022}. The ASAS-SN Variable Stars Database\footnote{\url{https://asas-sn.osu.edu/variables}} has revealed $\sim$ 688,000 variable stars, $\sim$ 239,000 of which are newly discovered. 
		
		The ZTF project is a new time-domain survey that can monitor the entire northern visible sky every three nights with the limiting magnitudes of \emph{g} $\sim$ 20.8 and \emph{r} $\sim$ 20.6 mag \citep{Masci2019}. Stellar variability is one of the main science goals of the ZTF project \citep{Graham2019}. There are more than 1.4 billion light curves in ZTF Data Release 2 ( ZTF DR2), and the ratio of the number of light curves with $\ge$ 20 epochs to the number of all light curves is $\sim$ 44\% in both the \emph{r} and \emph{g} bands. \cite{Chen2020} found that the majority of stars located in the northern Galactic plane contained $\sim$ 150 epochs of observations.
		
		The CNEOST is located at XuYi station of the Purple Mountain Observatory (PMO). The observations were carried out in five filters: \emph{CL} (no filter), \emph{VR} (\emph{V}+\emph{R}), \emph{SG} (SDSS-\emph{g}), \emph{SR} (SDSS-\emph{r}), and \emph{SI} (SDSS-\emph{i}). The \emph{V}-band limiting magnitude reaches $\sim$ 21 mag. The CNEOST has been in operation since 2006 and has generated data of 6.6 TB, totaling over 8.5 billion data records. 
		
		The WFST, jointly developed by the University of Science and Technology of China and the Purple Mountain Observatory, will be an image survey telescope and the most powerful time-domain survey equipment in the northern sky. This machine will be characterized by a 2.5-meter primary mirror and a prime-focus camera of a field of view (FOV) 7 square degrees filled with a 9 $\times$ 9K $\times$ 9K mosaic CCD detector. The images will be hold in the \emph{u}, \emph{g}, \emph{r}, \emph{i}, \emph{z} and \emph{w} bands and provide high-precision astrometric and photometric catalogs of objects down to \emph{r} $<$ 22.5 mag with a single exposure of 25 seconds, allowing us to precisely map structures of the Milky Way and the nearby universe. The WFST website\footnote{\url{http://wfst.ustc.edu.cn/main.htm}} shows the specific configuration and design parameters of the telescope. Currently, the science book is still in preparation, and we look forward to WFST being put into operation as soon as possible.
		
		In recent years, machine learning and deep learning have been widely used in variable star classification. \cite{KimBailer-Jones2016} constructed a random forest classifier based on the light curves of periodic variables from the OGLE and EROS-2 \citep{Tisserand2007} databases, which can automatically classify light curves into one of seven superclasses. They kindly provided a Python package “UPSILoN” containing the pre-trained random forest classifier that was later used for variable star classification by several scholars \citep{Jayasinghe2018a,Jayasinghe2018c,Kains2019,Hosenie2019}. \cite{Jayasinghe2019b} used a random forest classifier to analyse the ASAS-SN \emph{V}-band light curves of $\sim$ 412,000 known variable stars. In order to extend the census of RR Lyrae stars to highly reddened low-latitude regions of the central Milky Way, \cite{Dekany&Grebel2020} trained a deep long short-term memory recurrent neural network model \citep{Cho2014,Chung2014} based on the data from the VISTA Variables in the Vía Láctea \citep[VVV;][]{Dekany2018}. \cite{Kim2021} developed a single-band light curve classifier based on deep neural networks using OGLE and EROS-2 data and used transfer learning to solve the problem of training data scarcity by conveying knowledge from one data set to another. Again, they kindly provided a Python package “UPSILoN-T” that contained the pre-trained neural network and could optimize the model using a small set of light curves from the target survey. \cite{Sanchez-Saez2021} used a two-level random forest classifier to classify the sources from the ZTF database into 15 subclasses. The first level consisted of a single classifier that separated the sources as transient, stochastic, or periodic. The second-layer classifier consisted of three subclassifiers constructed using transient samples, periodic samples and stochastic samples respectively. Each subclassifier further divided the corresponding superclass in the first level into 4-6 subclasses. \cite{Christy2022} identified 1.5 $\times$ 10$^{6}$ variable star candidates using a random forest classifier from an input source list of $\sim$ 55 million isolated sources with \emph{g} $<$ 18 mag. 
		
		Along with the rapid development of sky survey facilities, the information we obtained about variables also improves rapidly. At present, there are many classification models, each based on its own set of observational features and applicable conditions. In order to make full use of the massive data from all kinds of survey programs, improve our understanding of the variables, and establish a variable star identification and classification tool for WFST, we propose a new period determination method and construct a random forest classifier. In Section~\ref{sec:packages2}, we describe the variable star data used to develop the period determination method and construct the random forest classifier. In Section~\ref{sec:packages3}, we introduce the period determination method. In Section~\ref{sec:packages4}, we construct a random forest classifier and show the classification performance of the classifier. In Section~\ref{sec:packages5}, we present the period accuracy and classification accuracy compared to the results of \cite{Chen2020} for ZTF DR2. In Section~\ref{sec:packages6}, we summarize the efficiency, advantages and disadvantages of the period determination method, the performance of the random forest classifier, and
		prospects for future work.
		
		\section{PHOTOMETRIC DATA} \label{sec:packages2}
		
		In order to develop a new period determination and an accurate classifier of variables, we build a variable sample based on the datasets from ASAS-SN and OGLE. We select variables with good data quality based on the \emph{classification probability} and \emph{lksl\_statistic}. The definitions of \emph{lksl\_statistic} \citep[Laflfler–Kinmann string length statistic;][]{Lafler&Kinman1965,Clarke2002} of periodic variables are shown in formula (1), where \emph{m$_{i}$} are the magnitudes sorted by ascending phase, \emph{$\bar{m}$} is the mean magnitude, \emph{N} is the number of measurements in the light curve. We examine different types of variables in the ASAS-SN Variable Star Database and find that variables with \emph{lksl\_statistic} $\leq$ 0.5 generally have less dispersion in phased light curves, higher \emph{classification probability}, and more accuracy periods. \cite{Jayasinghe2019b} used \emph{lksl\_statistic} $<$ 0.5 to count the number of light curves with good data quality. 
		\begin{equation}
			T\left(\phi|P\right)= \frac{\sum_{i=1}^{N}\left(m_{i+1} - m_{i}\right)^2}{\sum_{i=1}^{N}\left(m_{i} - \bar{m}\right)^2}\times\frac{\left(N-1\right)}{2}
		\end{equation}
		
		We select 220,907 \emph{V}-band periodic variables with \emph{classification probability} $\ge$ 0.6 (93.5\% samples with \emph{classification probability} $\ge$ 0.9) and \emph{lksl\_statistic} $\leq$ 0.5 from the ASAS-SN Variable Star Database \citep[The ASAS-SN Catalog of Variable Stars I - IX: ][]{Jayasinghe2018c,Jayasinghe2019a,Jayasinghe2019b,Jayasinghe2020a,Jayasinghe2021}. We also adopt the name scheme of ASAS-SN. Because the numbers of the Cepheids, RRD and ELL variables meeting the criteria are relatively small in ASAS-SN database, we select 20,247 \emph{I}-band periodic variables with \emph{lksl\_statistic} $\leq$ 0.5 from the OGLE database \citep{Soszynski2016a,Soszynski2016b,Soszynski2017a,Soszynski2017b,Soszynski2018,Soszynski2019,Soszynski2020,Udalski2018,Pawlak2016} and make the size of the sample to 241,154. Due to different name schemes between ASAS-SN and OGLE, we adopt the matching scheme in Table~\ref{tab:table1} to convert variability types. The number of variables of each class is shown in Table~\ref{tab:table2}. 
		
		\begin{table}
			\centering
			\caption{Variability types matching scheme between OGLE and ASAS-SN catalogs.}
			\label{tab:table1}
				\begin{tabular}{lccr} 
					\hline
					Superclass & OGLE subclass & ASAS-SN subclass\\
					\hline
					Type I Cepheids\\
					& F &  \\
					& F1 & DCEP\\
					& F12 & \\
					\cline{2-3}
					& 1 &  \\
					& 12 & \\
					& 13 & \\
					& 123 &  DCEPS\\
					& 2 & \\
					& 23 & \\
					\hline
					Type II Cepheids\\
					& BLHer & \multirow{4}{*}{CWB}\\
					& BLHer\_1O & \\
					& pWVir (period $<$ 8 d) &  \\
					& WVir (period $<$ 8 d) &  \\
					\cline{2-3}
					& pWVir (period $>$ 8 d) & \multirow{2}{*}{CWA} \\
					& WVir (period $>$ 8 d) &  \\
					\cline{2-3}
					& RVTau & RVA\\
					\hline
					Eclipsing binaries\\
					& ELL & ELL\\
					\hline
					RR Lyrae\\
					& RRd & RRD\\
					\hline
				\end{tabular}
		\end{table}
		
		\begin{table}
			\centering
			\caption{The number of sources in each class from ASAS-SN and OGLE databases.}
			\label{tab:table2}
			\begin{tabular}{lcccr} 
				\hline
				Superclass & Subclass & ASAS-SN & OGLE & Total\\
				\hline
				Cepheids\\
				& CWA & 446 & 430 & 876\\
				& CWB & 371 & 463 & 834\\
				& DCEP & 1306 & 4131 & 5437\\
				& DCEPS & 430 & 2715& 3145\\
				& RVA & 58 & 261 & 319\\
				Delta Scuti\\
				& DSCT & 1916 & 0 & 1916\\
				& HADS & 3144 & 0 & 3144\\
				Eclipsing binaries\\
				& EA & 37456 & 0 & 37456\\
				& EB & 20242 & 0 & 20242\\
				& ELL & 25 & 11896 & 11921\\
				& EW & 58898 & 0 & 58898\\
				Mira\\
				& M & 9162 & 0 & 9162\\
				Rotational variables\\
				& ROT & 10049 & 0 & 10049\\
				RR Lyrae\\
				& RRAB & 25671 & 0 & 25671\\
				& RRC & 8307 & 0 & 8307\\
				& RRD & 254 & 351 & 605\\
				Semiregular variables\\
				& SR & 43172 & 0 & 43172\\
				Irregular variables\\
				& L & 48682 & 0 & 48682\\
				& YSO & 1324 & 0 & 1324\\
				\hline
				Total &  & 270913 & 20247 & 291160\\
				\hline
			\end{tabular}
		\end{table}
		
		For every star in the sample, we extract 11 parameters (\emph{logP}, \emph{R$_{21}$}, \emph{R$_{31}$}, \emph{R$_{41}$}, $\sigma$, \emph{ALH}, \emph{M$_{s}$}, \emph{M$_{k}$}, \emph{MAD}, \emph{IQR}, \emph{A}) from each light curve and cross-match with  from the Gaia Early DR3 \citep{Bailer-Jones2021}, 2MASS \citep{Skrutskie2006}, and ALLWISE \citep{Cutri2013,Wright2010} catalogs to obtain additional 4 parameters (\emph{G$_{BP}$} - \emph{G$_{RP}$}, \emph{J} - \emph{K$_{s}$}, \emph{J} - \emph{H}, \emph{W1} $-$ \emph{W2}). The matching radii are 5.0 arcsec, 10.0 arcsec, and 10.0 arcsec, respectively. Moreover, we calculate 2 absolute Wensenheit magnitudes \emph{W$_{RP}$} and \emph{W$_{JK}$}  \citep{Madore1982,Lebzelter2018} for every periodic variable star. Object distance is necessary to calculate the two absolute magnitudes. Two distance evaluations (\texttt{r\_med\_photogeo}, \texttt{r\_med\_geo}) are given in Gaia Early DR3, and \texttt{r\_med\_photogeo} is preferred. The descriptions and definitions of these parameters are adopted from \cite{Jayasinghe2019b} and are shown in Table~\ref{tab:table3}. The \emph{classification probability}, \emph{lksl\_statistic}, period, object distance, \emph{G$_{BP}$} - \emph{G$_{RP}$}, \emph{J} - \emph{K$_{s}$}, \emph{J} - \emph{H}, \emph{W1} $-$ \emph{W2},  \emph{R$_{21}$}, \emph{R$_{31}$}, \emph{R$_{41}$}, \emph{W$_{RP}$}, and \emph{W$_{JK}$} of each periodic variable star can be directly or indirectly obtained from the ASAS-SN catalog. We use these parameter values and calculate remaining 7 parameter values. We use the periods of variables in the OGLE catalog to derive the \emph{lksl\_statistic} according to formula (1), \emph{R$_{21}$}, \emph{R$_{31}$}, \emph{R$_{41}$} and  calculate remaining 13 parameter values. 
		
		To test the efficiency of the method and classifier, we use irregular variables in ASAS-SN Variable Star Database and photometric data of periodic variables from ZTF DR2 sorted out by \cite{Chen2020}. We will separately present the photometric data of irregular variables from ASAS-SN and periodic variables from ZTF DR2 in Section~\ref{sec:packages4.3} and in Section~\ref{sec:packages5}, respectively.
		
		\begin{table*}
			\centering
			\caption{The 17 variability features used to train the random forest classifier and their importance.}
			\resizebox{\textwidth}{40mm}{
				\begin{tabular}{lccc} 
					\hline
					Feature & Description & Importance (\%) & Reference\\
					\hline
					\emph{logP} & Base 10 logarithm of the period & 24.2 &	-\\
					\emph{R$_{21}$} & Ratio between the amplitudes of the 2nd and 1st harmonics & 10.4	& -\\
					&   of the 6th-order Fourier model &  & \\
					\emph{$\sigma$} & Standard deviation of magnitude distribution & 8.1 & -\\
					\emph{A} &	Amplitude of the light curve (between the 5th and 95th percentiles)	& 7.0 &	-\\
					\emph{W$_{RP}$} & Absolute Wesenheit magnitude in Gaia Early DR3 \emph{G$_{RP}$}-band & 6.1 & \cite{Bailer-Jones2021}, \cite{Skrutskie2006},\\
					&  &  & \cite{Madore1982}\\
					\emph{M$_{s}$} & Skewness of the magnitude distribution	& 5.5 & -\\
					\emph{IQR}	& Difference between the 75th and 25th percentiles in magnitude	& 5.4 &	-\\
					\emph{M$_{k}$} & Kurtosis of the magnitude distribution	& 4.8 & -\\	
					\emph{W$_{JK}$} & Absolute Wesenheit magnitude in 2MASS \emph{K$_{s}$}-band & 4.4 & \cite{Bailer-Jones2021}, \cite{Skrutskie2006},\\
					&  &  &  \cite{Madore1982}\\
					\emph{MAD} & Median absolute deviation of magnitude distribution & 3.8 & -\\
					\emph{R$_{41}$} & Ratio between the amplitudes of the 4th and 1st harmonics & 3.7 & -\\
					&   of the 6th-order Fourier model &  & \\
					\emph{G$_{BP}$} $-$ \emph{G$_{RP}$} & Gaia Early DR3 \emph{G$_{BP}$} $-$ \emph{G$_{RP}$} color & 3.1	& \cite{GaiaCollaboration2021}\\
					\emph{J} $-$ \emph{H} & 2MASS \emph{J} $-$ \emph{H} color & 3.0 & \cite{Skrutskie2006}\\
					\emph{R$_{31}$} & Ratio between the amplitudes of the 3rd and 1st harmonics & 2.9 & -\\
					&   of the 6th-order Fourier model &  & \\
					\emph{J} $-$ \emph{K$_{s}$} & 2MASS \emph{J} $-$ \emph{K$_{s}$} color & 2.7 & \cite{Skrutskie2006}\\
					\emph{W1} $-$ \emph{W2} & WISE \emph{W1} $-$ \emph{W2} color & 2.3 & \cite{Cutri2013}, \cite{Wright2010}\\
					\emph{ALH}	& Ratio of magnitudes fainter or brighter than the average  & 2.3 & \cite{KimBailer-Jones2016}\\
					\hline
				\end{tabular}
			}
			\label{tab:table3}
		\end{table*}

		\section{A NEW PERIOD DETERMINATION METHOD}
		\label{sec:packages3}
		
		\subsection{Commonly used period-finding algorithms}
		\label{sec:packages3.1}
		Period is a very important property of periodic variable stars. The period-luminosity relationships can help to estimate the luminosity and distance of variable stars. The variability types and evolution stages of variable stars can be roughly inferred from their periods. At present, the commonly used period determination methods are Generalized Lomb–Scargle \citep[GLS;][]{Scargle1982,Zechmeister&Kurster2009}, Box Least Squares \citep[BLS;][]{Kovacs&Zucker&Mazeh2002}, Multi-Harmonic Analysis Of Variance \citep[MHAOV;][]{Schwarzenberg-Czerny1996}, and Phase Dispersion Minimum \citep[PDM;][]{Stellingwerf1978}. 
		
		The GLS algorithm is good at deriving periods for variables with sinusoidal-like variability in phased light curves. However, for eclipsing binaries, the periods derived by the GLS algorithm have values that are half of the true periods. \cite{Graham2013} suggested that the best approach was to double the period once the light curve was confirmed to be that of an eclipsing binary. 
		
		The BLS algorithm is designed to look for planetary transits. Because of the similarities in light curves between planetary transits and stellar transits, the BLS algorithm is more suitable to derive periods for eclipsing binaries, but it may take a long time to find periods. The efficiency of the MHAOV algorithm in deriving accurate periods for non-sinusoidal variability is similar to that of the BLS algorithm, but the two algorithm are not good at deriving periods for long-period variables. The PDM algorithm is useful to find periods for variables with long time-span observations and non-sinusoidal variability and works well for light curves with fewer photometry data. However, the PDM algorithm is a traversal algorithm, which is very time-consuming.  \cite{Jayasinghe2018c,Jayasinghe2019a,Jayasinghe2019b} combined multiple algorithms and then selected the periods with the highest score calculated by the SuperSmoother function \citep{Reimann1994,Vanderplas2016} as the best periods, where the score is the normalized power of the candidate period in the periodogram calculated by the SuperSmoother algorithm from the observation data.
		
		\texttt{Vartools} \citep{Hartman&Bakos2016} is an open source command-line utility, written in \texttt{C}, for analysing astronomical time-series data, especially light curves. The program provides many period-finding algorithms such as the GLS periodogram, the BLS transit search routine, the Analysis of Variance periodogram, the Discrete Fourier Transform including the CLEAN algorithm, and the Weighted Wavelet Z-Transform. These algorithms can quickly find periods, but the command-line utility is not as flexible and convenient as \texttt{PYTHON} and \texttt{MATLAB} programs. \cite{Pawlak2019} used the GLS and BLS algorithms in the \texttt{VARTOOLS} and selected the candidates for periodic variables based on the signal-to-noise ratio (SNR) determined within the periodograms of both methods. 
		
		We used the method described in \cite{Jayasinghe2018c,Jayasinghe2019a,Jayasinghe2019b} to measure the periods of periodic variables. Comparing to using only the GLS algorithm, the period accuracy of eclipsing binaries has improved, but there are still many eclipsing binaries with inaccurate periods that require subsequent adjustments and the period accuracy of variables with sinusoidal-like variability is reduced. At the same time, because of the inclusion of the BLS and MHAOV algorithms, the period determination process usually takes a long time. 
		
		\subsection{Separation method for eclipsing binaries and NEB variables}
		\label{sec:packages3.2}
		As described above, it is difficult to derive accurate periods for different variables by using only one algorithm or combining several algorithms in a single step involving simple criteria. Since the efficiency of the GLS algorithm
		is significantly affected by eclipsing binaries, we can first divide different variables into eclipsing binaries and NEB variables (other types of periodic variable stars other than eclipsing binaries), and then use the GLS algorithm to derive the periods. If these variables are in eclipsing binaries, the periods are doubled. If they are in NEB variables, the periods are unchanged. Through a large number of tests, we finally find five parameters (\emph{ALH}, \emph{M$_{k}$}, $\sigma$, \emph{G$_{BP}$} - \emph{G$_{RP}$}, \emph{J} - \emph{H}) that can be used to separate eclipsing binaries and NEB variables. The five parameters and the period are both obtained from photometric measurements. The period is mainly obtained from the time domain information, but the five parameters are calculated from the magnitude and color information. Therefore, the parameters used in our separation method are completely independent of the period.
		
\begin{figure}
			\centering
			\includegraphics[width=\textwidth]{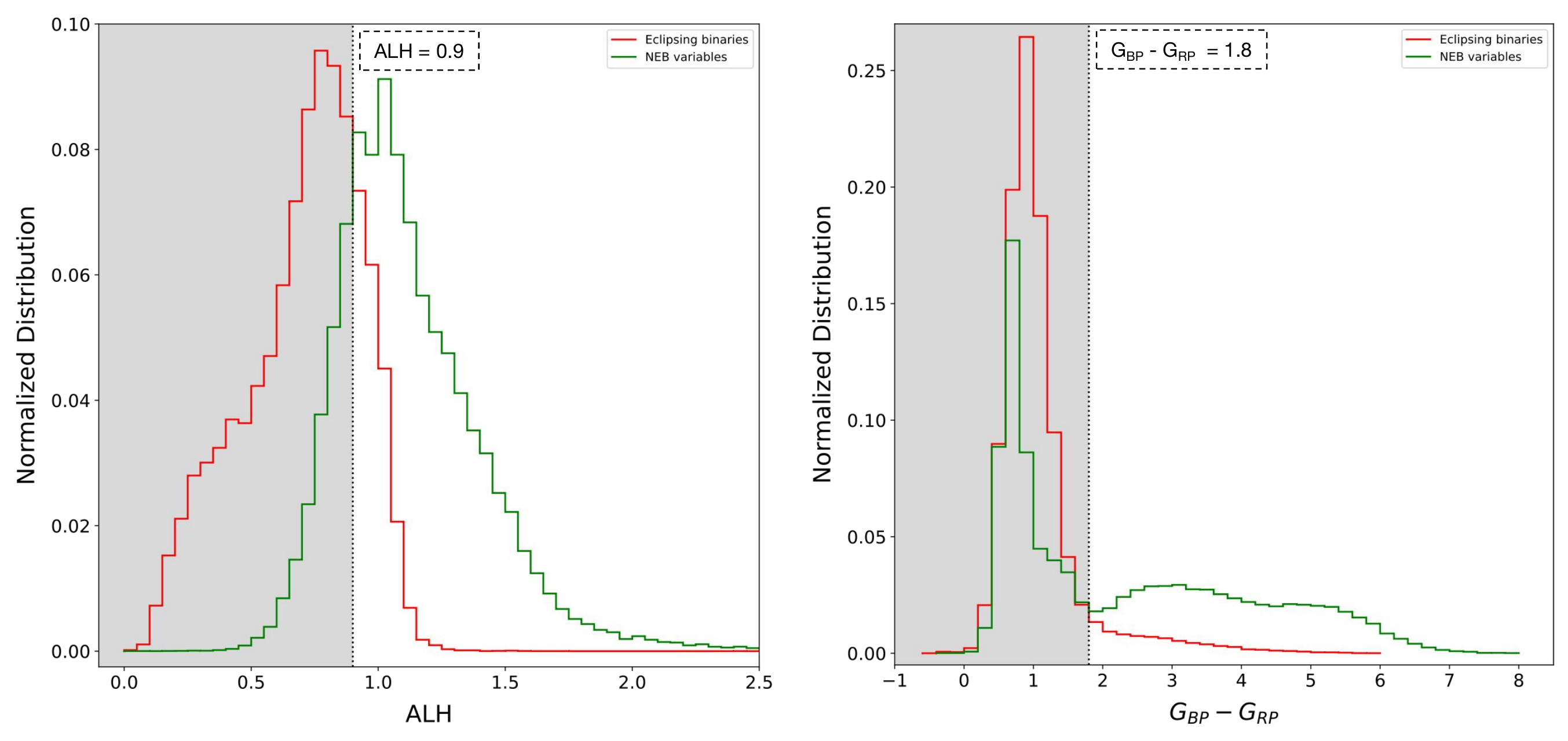}
			\caption{The distributions of all eclipsing binaries and NEB variables in \emph{ALH} and \emph{G$_{BP}$} - \emph{G$_{RP}$} diagrams. The region meeting \protect\hyperlink{criterion1}{criterion 1} is shaded in gray.}
			\label{fig:figure1}
		\end{figure}

\begin{figure}
			\centering
			\includegraphics[width=\textwidth]{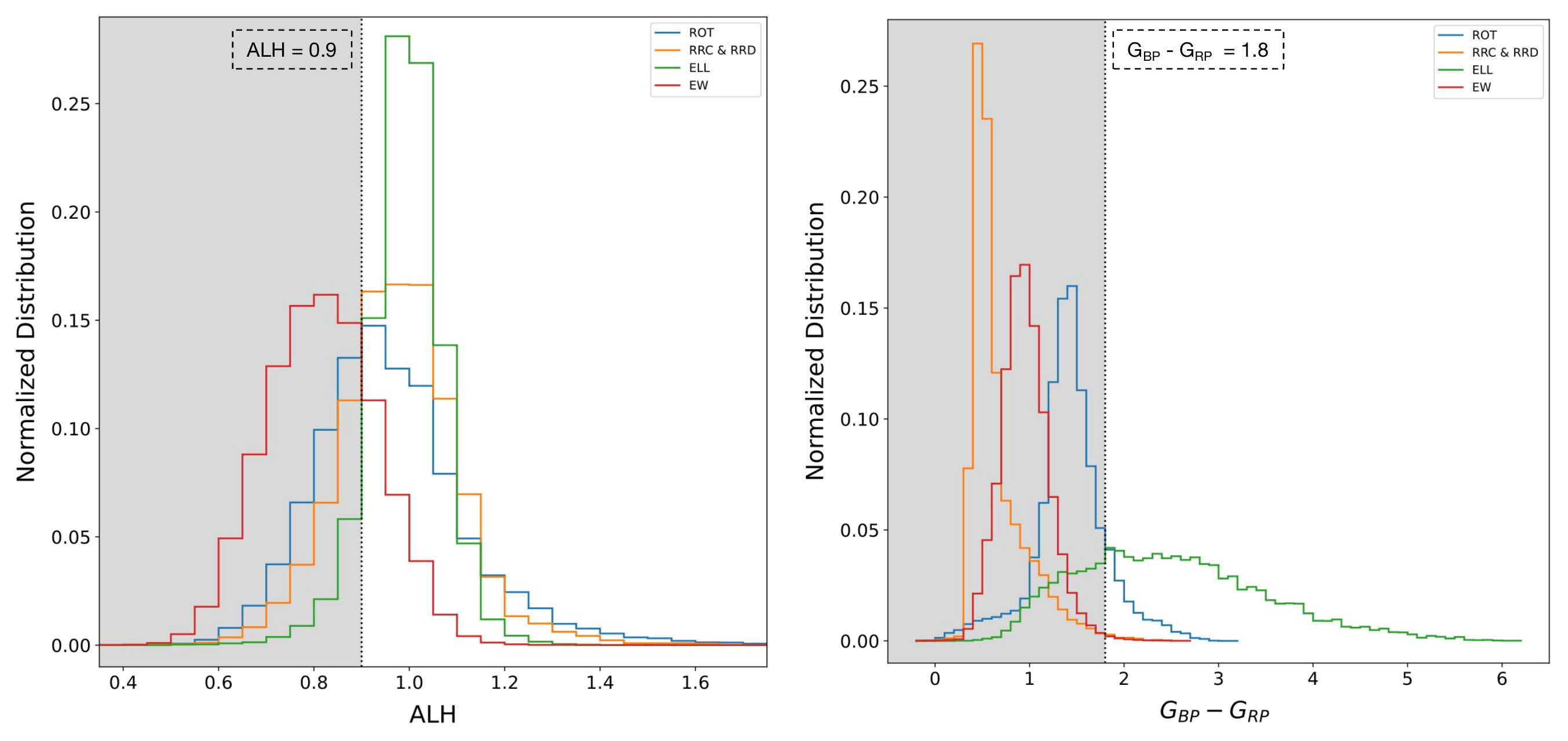}
			\caption{The distributions of all ROT, RRC\&RRD, ELL, EW variables in \emph{ALH} and \emph{G$_{BP}$} - \emph{G$_{RP}$} diagrams. The region meeting \protect\hyperlink{criterion1}{criterion 1} is shaded in gray.}
			\label{fig:figure2}
		\end{figure}

		We find that most of the eclipsing binaries and NEB variables can be separated based on the \emph{ALH} and \emph{G$_{BP}$} - \emph{G$_{RP}$}. As shown in Fig.~\ref{fig:figure1}, the distributions between eclipsing binaries and NEB variables are evident in the \emph{ALH} diagram. After strengthening the restriction of \emph{G$_{BP}$} - \emph{G$_{RP}$} $\leq$ 1.8, it can separate nearly 100\% of long-period variable stars (Miras and SRs) to be separated from eclipsing binaries. The variable stars meeting criterion 1 (\hypertarget{criterion1}{\emph{ALH} $\leq$ 0.9 and \emph{G$_{BP}$} - \emph{G$_{RP}$} $\leq$ 1.8}) are eclipsing binaries with high probability. The variable stars meeting the opposite criterion (i.e. criterion 2: \hypertarget{criterion2}{\emph{ALH} $>$ 0.9 or \emph{G$_{BP}$} - \emph{G$_{RP}$} $>$ 1.8}) are NEB variables with high probability. However, $\sim$ 31\% of ROT and $\sim$ 25\% of RRC\&RRD variables meet \hyperlink{criterion1}{criterion 1} and are mis-classified as eclipsing binaries. Since the number of RRD variables is small, RRD and RRC variables are considered here as the same type during the separation process. It does not cause significant changes in the separation results. Their differences are discussed in the later sections. $\sim$ 96\% of ELL and $\sim$ 24\% of EW variables meet \hyperlink{criterion2}{criterion 2} and are mis-classified as NEB variables. Fig.~\ref{fig:figure2} shows the mixed phenomena.
		
\begin{figure}
			\centering
			\includegraphics[width=\textwidth]{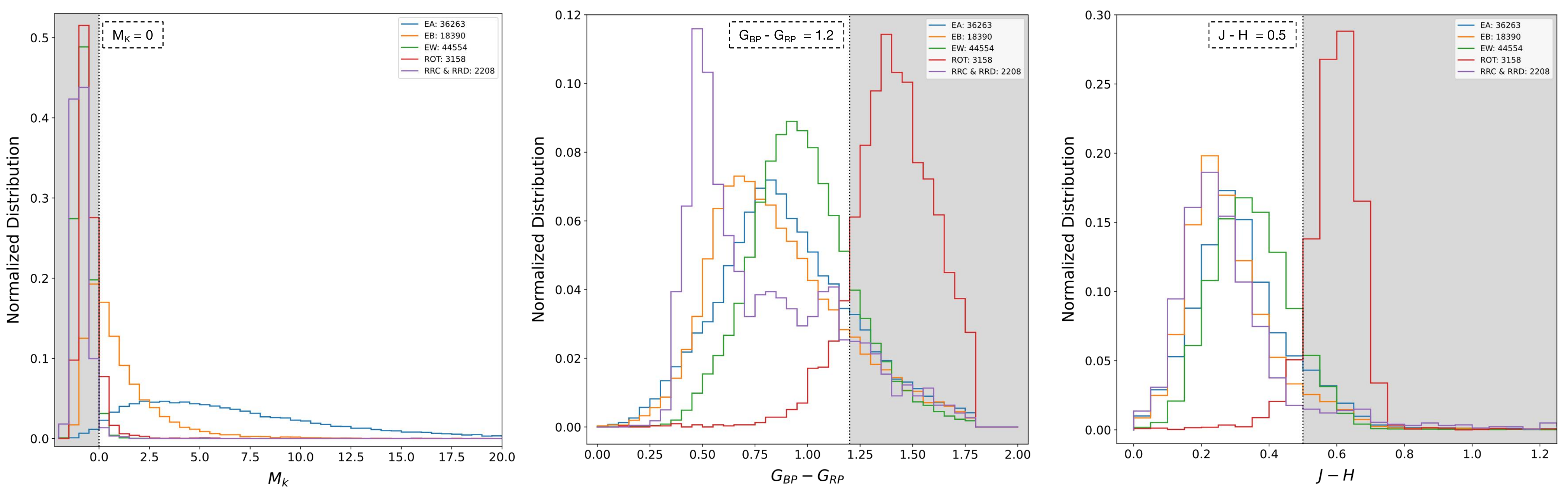}
			\caption{The distributions of EA, EB, EW, ROT and RRC\&RRD variables in \emph{M$_{k}$}, \emph{G$_{BP}$} - \emph{G$_{RP}$} and \emph{J - H} diagrams. The numbers of RRC\&RRD and ROT variables meeting \protect\hyperlink{criterion1}{criterion 1} are shown in the legend. The region meeting \protect\hyperlink{criterion1-2}{criterion 1-2} is shaded in gray.}
			\label{fig:figure3}
		\end{figure}

		It is possible to reduce these mis-classifications by using additional information. We find that some ROT variables can be separated from eclipsing binaries based on the \emph{M$_{k}$}, \emph{G$_{BP}$} - \emph{G$_{RP}$} and \emph{J - H}. As shown in Fig.~\ref{fig:figure3}, many ROT and RRC\&RRD variables meeting \hyperlink{criterion1}{criterion 1} are in the gray region of the \emph{M$_{k}$} diagram, but so are the EW and EB variables. In order to separate as many ROT and RRC\&RRD variables as possible and not to reduce the purity of EB and EW variables, we add the restrictions of \emph{G$_{BP}$} - \emph{G$_{RP}$} $\ge$ 1.2 and \emph{J - H} $\ge$ 0.5. The variables meeting criterion 1-1 (\hypertarget{criterion1-1}{\emph{M$_{k}$} $>$ 0 or \emph{G$_{BP}$} - \emph{G$_{RP}$} $<$ 1.2 or \emph{J - H} $<$ 0.5}) are considered the ture eclipsing binaries. The variables meeting the opposite criterion (i.e. criterion 1-2: \hypertarget{criterion1-2}{\emph{M$_{k}$} $\le$ 0 and \emph{G$_{BP}$} - \emph{G$_{RP}$} $\ge$ 1.2 and \emph{J - H} $\ge$ 0.5}) are considered as NEB variables. About 24\% of ROT variables meet \hyperlink{criterion1-2}{criterion 1-2}, these ROT variables can be selected from eclipsing binaries. Eventually, only $\sim$ 7\% of ROT variables are mis-classified as eclipsing binaries. However, $\sim$ 23\% of RRC\&RRD variables exactly meet \hyperlink{criterion1-1}{criterion 1-1} and cannot be separated from eclipsing binaries. Unfortunately, $\sim$ 5\% of EW variables meet \hyperlink{criterion1-2}{criterion 1-2}, these varibales are mis-classified as NEB variables. 
		
\begin{figure}
			\centering
			\includegraphics[width=\textwidth]{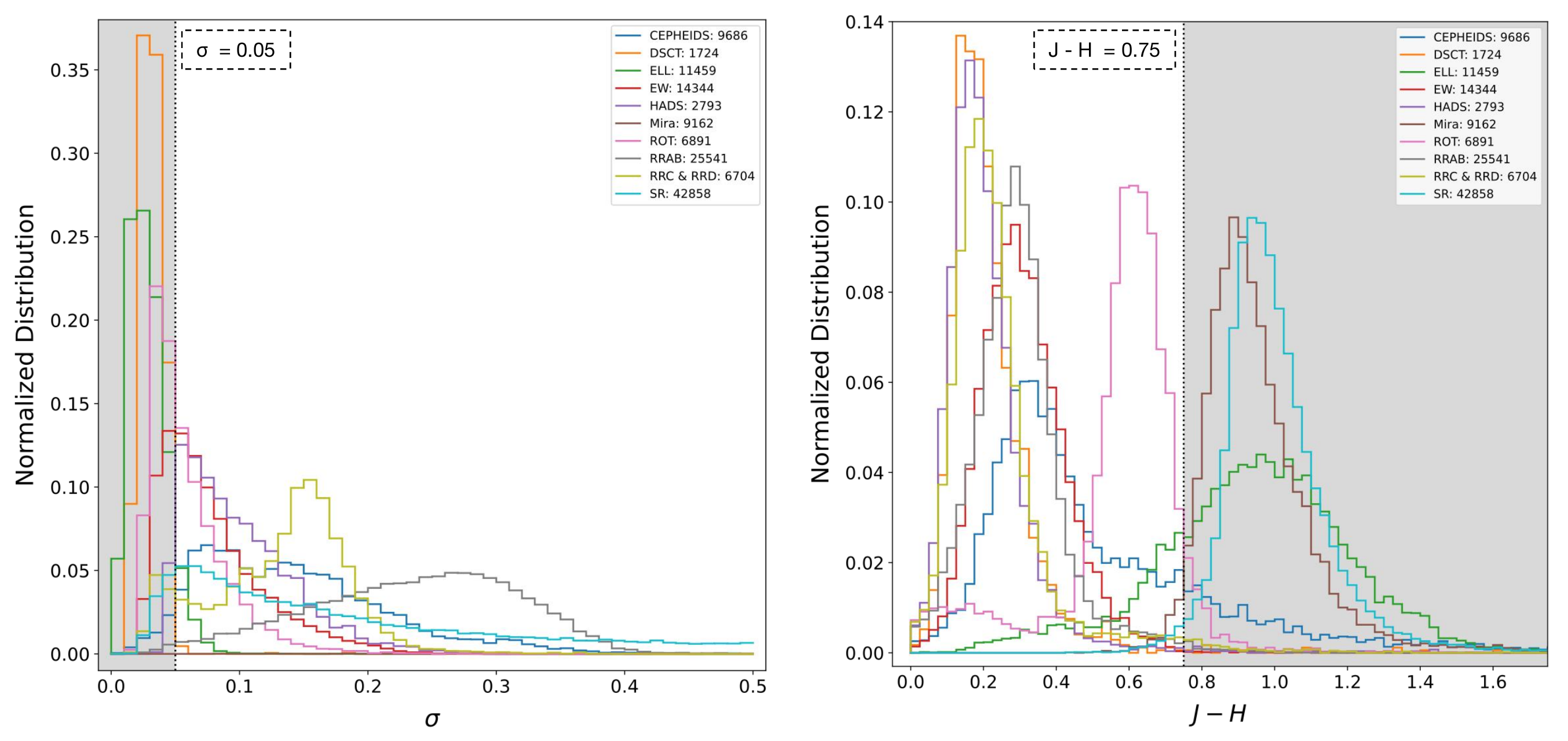}
			\caption{The distributions of ELL and EW variables and NEB variables in \emph{$\sigma$} and \emph{J - H} diagrams. To show more details, Miras' distribution is not fully shown in the \emph{$\sigma$} diagram. The numbers of  variables meeting \protect\hyperlink{criterion2}{criterion 2} are shown in the legend. The region meeting  \protect\hyperlink{criterion2-1}{criterion 2-1} is shaded in gray.}
			\label{fig:figure4}
		\end{figure}

		We also find that many ELL variables can be separated from NEB variables based on the \emph{$\sigma$} and \emph{J} - \emph{H}.  As shown in Fig.~\ref{fig:figure4}, many ELL variables meeting \hyperlink{criterion2}{criterion 2} are in the gray region of the \emph{$\sigma$} diagram, but so are the DSCT and ROT variables. In order to separate as many ELL variables as possible and not to reduce the purity of DSCT and ROT variables, we add the restriction of \emph{J - H} $\ge$ 0.75 on these variables. The variables meeting criterion 2-1 (\hypertarget{criterion2-1}{\emph{$\sigma$} $\leq$ 0.05 and \emph{J} - \emph{H} $\ge$ 0.75}) are considered as eclipsing binaries. The variables meeting the opposite criterion (i.e. criterion 2-2: \hypertarget{criterion2-2}{\emph{$\sigma$} $>$ 0.05 or \emph{J} - \emph{H} $<$ 0.75}) are considered as the ture NEB variables. About 70\% of ELL variables meet  \hyperlink{criterion2-1}{criterion 2-1}, these ELL variables can be selected from NEB variables. Eventually, $\sim$ 26\% of ELL variables are mis-classified as NEB variables. However, $\sim$ 24\% of EW variables exactly meet \hyperlink{criterion2-2}{criterion 2-2} and cannot be seperated from NEB variables, so $\sim$ 29\% of EW variables are mis-classified as NEB variables. 
		
		The proportions of different types of periodic variables meeting 6 criteria are shown in Table~\ref{tab:table4}. Because the slight influence of the four sub-criteria on the purity of different types of variable stars is not mentioned, the final results are $\pm$(1\%-2\%) different from the descriptions above. About 70\%-97\% of eclipsing binaries can be separated correctly and the periods of these variables need to be doubled. Except for RRC variables, more than 90\% of NEB variables can be separated correctly. The values in Table~\ref{tab:table4} and the distributions in Fig.~\ref{fig:figure1} -- Fig.~\ref{fig:figure4} indicate that the eclipsing binaries and NEB variables can be efficiently separated based on the 5 parameters (\emph{ALH}, \emph{M$_{k}$}, $\sigma$, \emph{G$_{BP}$} - \emph{G$_{RP}$}, \emph{J} - \emph{H}) and 6 criteria (\hyperlink{criterion1}{1}, \hyperlink{criterion1-1}{1-1}, \hyperlink{criterion1-2}{1-2}, \hyperlink{criterion2}{2},  \hyperlink{criterion2-1}{2-1},  \hyperlink{criterion2-2}{2-2}) and indirectly prove that our method can accurately derive the accurate periods for periodic variables.
		
		\begin{table}
			\caption{The percentage of meeting 6 criteria for periodic variable stars from ASAS-SN and OGLE.}
				\begin{tabular}{lccc} 
					\hline
					Superclass & Subclass & Meeting \protect\hyperlink{criterion1-1}{criteria 1-1,} \protect\hyperlink{criterion2-1}{2-1} & Meeting \protect\hyperlink{criterion1-2}{criteria 1-2,} \protect\hyperlink{criterion2-2}{2-2}\\
					\hline
					Eclipsing binaries\\
					& EA (37456) & 96.6\% & 3.4\%\\
					& EB (20242) & 90.0\% & 10.0\%\\
					& ELL (11921) & 71.5\% & 28.5\%\\
					& EW (58898) &70.4\% & 29.6\%\\
					NEB variables\\
					& Cepheids (10611) & 7.9\% & 92.1\%\\
					& Delta Scuti (5060) & 8.8\% & 91.2\%\\
					& Mira (9162) & 0.0\% & 100.0\%\\
					& ROT (10049) & 9.1\% & 90.9\%\\
					& RRAB (25671) & 0.6\% & 99.4\%\\
					& RRC (8307) & 24.7\% & 75.3\%\\
					& RRD (605) & 5.8\% & 94.2\%\\
					& RRC\&RRD (8912) & 23.4\% & 76.6\%\\
					& SR (43172) & 9.1\% & 90.9\%\\
					\hline
				\end{tabular}
			\label{tab:table4}
		\end{table}

		\subsection{Period search} \label{sec:packages3.3}
		Once eclipsing binaries and NEB variables are separated, we can find the best periods for each type of the variables by suitable algorithm.
		
		For each light curve,  we save 10 strongest frequency peaks from the GLS periodogram and consider them as 10 most possible periods. We import  \texttt{pgen\_lsp} from \texttt{astrobase.periodbase} to implement the GLS periodogram \citep{Bhatti2021}. The difference is that the adjacent frequency interval is greater than 0.0001 and 0.01 for eclipsing binaries (i.e. meeting \hyperlink{criterion1-1}{criterion 1-1} or \hyperlink{criterion2-1}{2-1}) and NEB variables (i.e. meeting \hyperlink{criterion1-2}{criterion 1-2} or \hyperlink{criterion2-2}{2-2} ), respectively.
		
		To remove false and weak period signals later, we save 10 strongest peaks with adjacent frequency intervals greater than 0.05 from the window function and consider them as 10 aliased periods. We import \texttt{LombScargle} from \texttt{astropy.timeseries} to implement the window function \citep{AstropyCollaboration2013}. Two flase alarm probability levels (FAP1 and FAP2) corresponding to 1 $\times$ 10$^{-2}$ and 1 $\times$ 10$^{-15}$ in this window function are also saved. 
		
\begin{figure*}
			\centering
			\includegraphics[width=\textwidth]{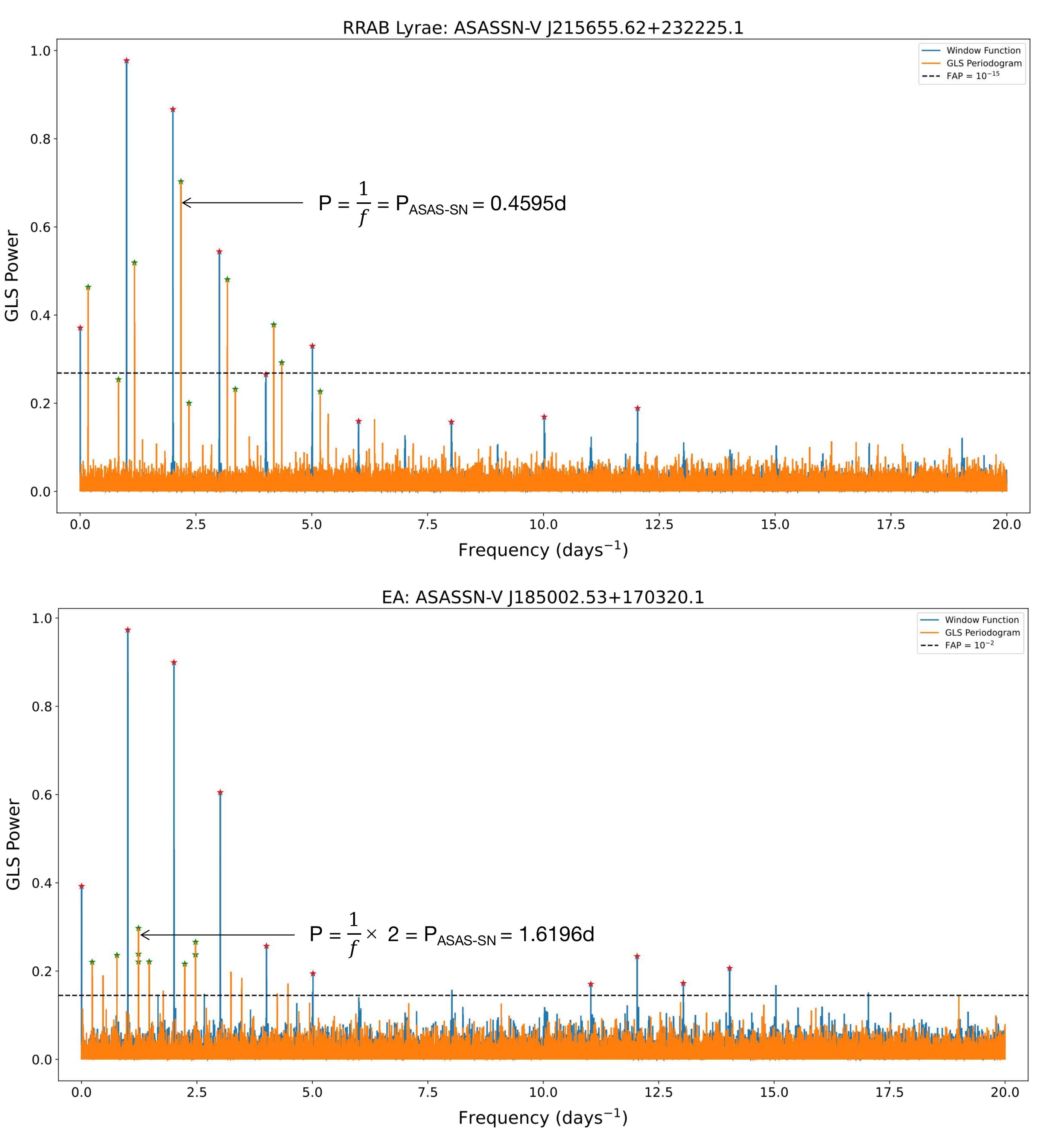}
			\caption{The window function and GLS periodogram of the RRAB variable (ASASSN-V J215655.62+232225.1) and the EA variable (ASASSN-V J185002.53+170320.1). The 10 strongest peaks of window function and GLS periodogram are marked in red pentagram and green pentagram. Two FAP levels shown in dotted lines help to exclud weak signals.}
			\label{fig:figure5}
		\end{figure*}

		For the eclipsing binaries, those periods within 10$^{-3}$ of an aliased period or with a FAP $>$ FAP1 are removed from consideration.  For the NEB variables, those periods within 10$^{-3}$ of an aliased period or with a FAP $>$ FAP2 are removed from consideration. We find that the signal strength of eclipsing binaries are usually weaker than that of NEB variables, especially EA variables, so we set two FAP levels to avoid a large number of eclipsing binaries being mis-removed in our work. The window function and GLS periodogram of a NEB variable star and an eclipsing binary are shown in  Fig.~\ref{fig:figure5}. 
		
		Moreover, the SuperSmoother function is used to calculate scores for the remaining possible periods, and the possible period with the highest score is considered to be the best matched period. In a rarely happening situation, one possible period may have a lower score but a higher power (i.e. a lower FAP) compared with the other possible periods. Still, the possible period with the higher score is selected as the final choice. We import \texttt{SuperSmoother} from \texttt{gatspy.periodic.supersmoother} to implement the SuperSmoother function \citep{Reimann1994,Vanderplas2016}. If the variable star is regarded as an eclipsing binaries (i.e. meeting \hyperlink{criterion1-1}{criterion 1-1} or \hyperlink{criterion2-1}{criterion 2-1}), the finally assigned period is twice the best matched period. If the variable star is regarded as a NEB variable star (i.e. meeting \hyperlink{criterion1-2}{criterion 1-2} or \hyperlink{criterion2-2}{criterion 2-2}), the finally assigned period is the same as the best matched period. 
		
\begin{figure*}
			\centering
			\includegraphics[width=0.9\textwidth,height=1.3\textwidth]{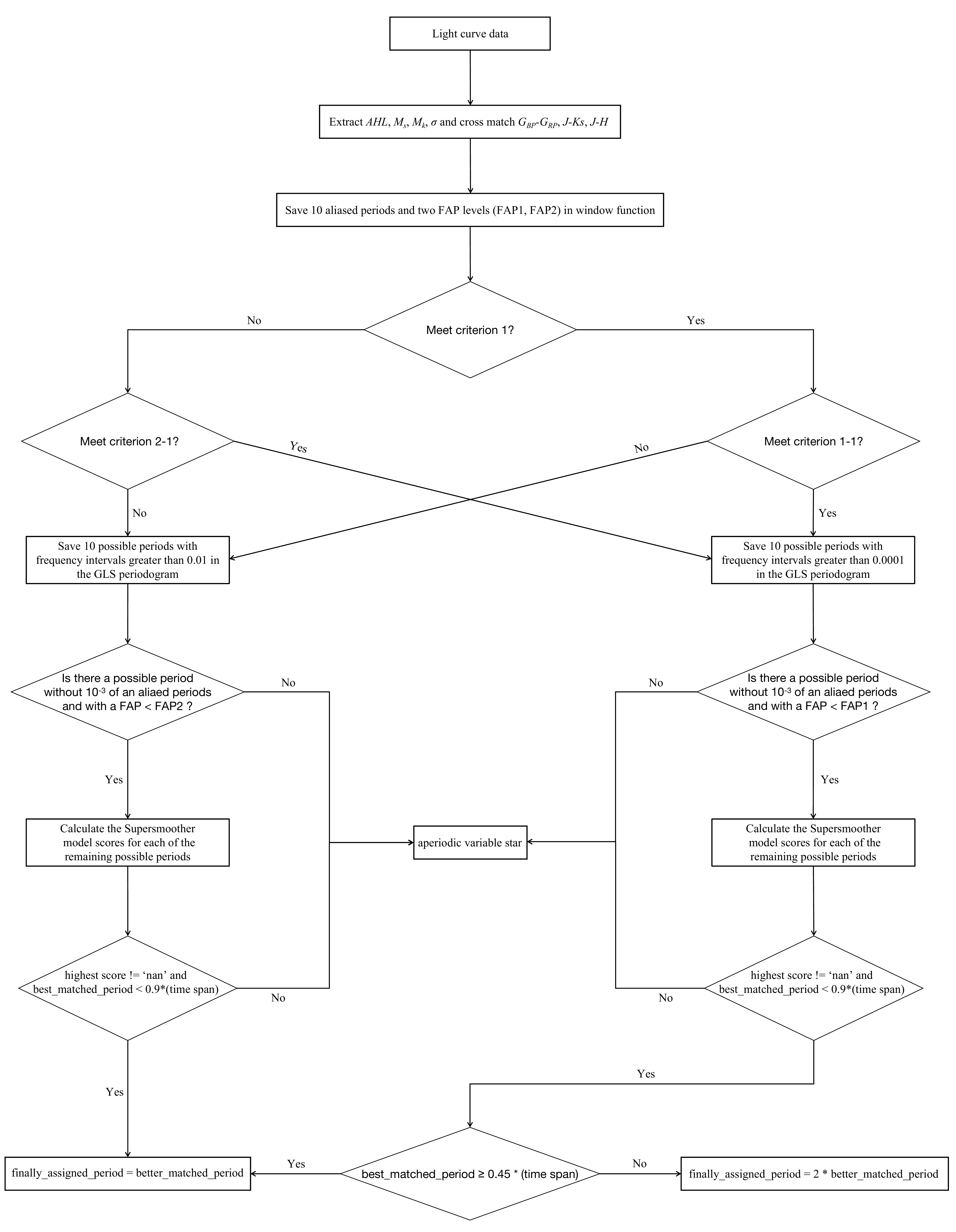}
			\caption{A schematic of the period determining procedure.}
			\label{fig:figure6}
		\end{figure*}

		However, if no possible period is left, the sample is considered an aperiodic variable star. It needs to be emphasized that the periods are searched over the range 0.05 d $\leq$ \emph{P} $\leq$ time span with the frequency step of 0.0001 day$^{-1}$ when we apply the GLS algorithm. Fig.~\ref{fig:figure6} shows a schematic of the complete period determining procedure.
		
		\subsection{Period accuracy comparison with ASAS-SN $\&$ OGLE}
		\label{sec:packages3.4}
		To demonstrate the efficiency of the new method, we select randomly 2,000 variables for each variability type (605 samples for RRD variables) from 241,154 variables, derive the periods using our method, and compare periods with those of the ASAS-SN and OGLE catalogs. The derived periods are considered the same as the original cataloged values if the differences are less than $\pm$5\%. In fact, for long-period variables such as Miras and SRs, the photometric data over a longer time span are necessary for determining the periods, the period deviations of long-period variables are greater than those of the short-period variables, and the period differences can be greater than $\pm$5\%. The comparison results are shown in Table~\ref{tab:table5}. We can see that 17.9\% of the EA variables and 18.8\% of the SR variables are considered as aperiodic variables, that may be related to the weak signal strength in the periodogram of EA variables or the characteritics of varying levels of periodicity of SR variables, but these variables are regarded as periodic variables in the ASAS-SN database. According to our method, there should be a close correlation between the values in Table~\ref{tab:table4} and Table~\ref{tab:table5}. Under the assumption that the periods in ASAS-SN and OGLE catalogs are 100\% correct, comparing the values in Table~\ref{tab:table4} with those in Table~\ref{tab:table5}, the proportions of EB, ELL, and EW variables meeting the \hyperlink{criterion1-1}{criteria 1-1} and \hyperlink{criterion2-1}{2-1} are similar to the period accuracy of these types of variables, but for the EA variables, the latter is about 15\% lower than the former. The EA, EB, EW, and ELL variables are all eclipsing binaries, but have different phased light curve characteristics. The EA variables may or may not have a small secondary eclipse, so the periods are easily affected by parameter values of the GLS algorithm and the number of measurements in the light curves. However, the period accuracy of EA variables also reaches 81\%, which indicates that the new method is very efficient for the period determination of EA variables.
		
		\begin{table}
			\caption{The period comparison between our own results and ASAS-SN, OGLE catalogs for periodic variable stars (the periods with differences less than $\pm$5\% are considered to be the same).}
			\centering
				\begin{tabular}{c|c|cc} 
					\hline
					ASAS-SN \& OGLE & \multicolumn{3}{c}{This work}\\
					\hline
					\multirow{2}{*}{Class} & \multirow{2}{*}{Aperiodic} & \multicolumn{2}{c}{Periodic}\\
					\cline{3-4}
					&  & same period & different period\\
					\hline
					Cepheids (2000) & 0.6\% & 92.5\% & 7.5\%\\
					Delta Scuti (2000) & 3.7\% & 89.3\% & 10.7\%\\
					EA (2000)  & 17.9\% & 80.9\% & 19.1\%\\
					EB (2000)  & 2.2\% & 89.1\% & 10.9\%\\
					ELL (2000) & 0.1\% & 72.1\% & 27.9\%\\
					EW (2000) & 0.8\% & 70.1\% & 29.9\%\\
					Mira (2000) & 4.2\% & 89.1\% & 10.9\%\\
					ROT (2000) & 8.0\% & 82.7\% & 17.3\%\\
					RRAB (2000) & 2.4\% & 99.0\% & 1.0\%\\
					RRC (2000) & 0.7\% & 74.7\% & 25.3\%\\
					RRD (605) & 5.0\% & 92.7\% & 7.3\%\\
					SR (2000) & 18.8\% & 75.0\% & 25.0\%\\
					\hline
				\end{tabular}
			\label{tab:table5}
		\end{table}
		
		The proportion values of Cepheids, Delta Scuti, ROT, RRAB, RRC, and RRD variables meeting \hyperlink{criterion1-2}{criteria 1-2} and \hyperlink{criterion2-2}{2-2} are similar to the period accuracy values of these types of variables, but for Mira and SR variables, the proportion values are about 11\% and 16\% lower than the period accuracy values. For long-period variable stars, the peaks with obvious periodicity on the periodogram tend to correspond to very small frequencies. Even very small frequency differences can have a large effect on the periods, so the periods with differences less than $\pm$(10\%-20\%) are very possible for Mira and SR variables. With differences less than $\pm$5\%, the period accuracy of Mira variables is as high as 89\%, and the period accuracy of SR variables is also up to 75\%, which indicates that the new method is also very efficient for period determination of Mira and SR variables. 
		
		We evaluate the merit of the 5\% difference criterion by computing the period consistencies under different criteria for short- and long-period variables. The results are listed in Table 6. It shows that the period differences of most types of short-period variables are less than 0.1\%. For ELL and ROT variables, the period differences are less than 1\%. For long-period variables, the period differences of 93\% Miras and 89\% SRs are less than 10\%. Therefore, the 5\% difference criterion is highly effective for short-period variables, but less effective for long-period variables.
		
		\begin{table}
			\caption{The period comparison between our own results and ASAS-SN, OGLE catalogs for periodic variable stars under other difference criteria.}
			\centering
				\begin{tabular}{c|ccccc} 
					\hline
					ASAS-SN \& OGLE & \multicolumn{5}{c}{This work}\\
					\hline
					\multirow{2}{*}{Short-period variables} & \multicolumn{5}{c}{Period accuracy}\\
					\cline{2-6}
					& $\pm$0.01\% & $\pm$0.1\% & $\pm$1\% & $\pm$3\% & $\pm$5\%\\
					\hline
					Cepheids (2000) & 41.5\% & 86.7\% & 92.5\% & 92.5\% & 92.5\%\\
					Delta Scuti (2000) & 89.3\% & 89.3\% & 89.3\% & 89.3\% & 89.3\%\\
					EA (2000)  & 70.2\% & 80.7\% & 80.8\% & 80.9\% & 80.9\%\\
					EB (2000)  & 85.7\% & 89.0\% & 89.1\% & 89.1\% & 89.1\%\\
					ELL (2000) & 20.5\% & 43.8\% & 71.6\% & 72.1\% & 72.1\%\\
					EW (2000) & 70.0\% & 70.1\% & 70.1\% & 70.1\% & 70.1\%\\
					ROT (2000) & 26.4\% & 67.0\% & 82.1\% & 82.7\% & 82.7\%\\
					RRAB (2000) & 98.6\% & 99.0\% & 99.0\% & 99.0\% & 99.0\%\\
					RRC(2000) & 74.5\% & 74.5\% & 74.5\% & 74.5\% & 74.5\%\\
					RRD (605) & 92.3\% & 92.7\% & 92.7\% & 92.7\% & 92.7\%\\
					\hline
					\multirow{2}{*}{Long-period variables} & \multicolumn{5}{c}{Period accuracy}\\
					\cline{2-6}
					& $\pm$1\% & $\pm$3\% & $\pm$5\% & $\pm$10\% & $\pm$20\%\\
					\hline
					Mira (2000) & 42.7\% & 79.2\% & 89.1\% & 93.1\% & 94.1\%\\
					SR (2000) & 37.8\% & 66.5\% & 75.0\% & 81.2\% & 83.4\%\\
					\hline
				\end{tabular}
			\label{tab:table6}
		\end{table}

		\section{RANDOM FOREST CLASSIFIER} \label{sec:packages4}
		
		\subsection{Training a random forest classifier}
		\label{sec:packages4.1}
		After deriving periods for periodic variables, we build a random forest classifier to classify different variables. In the past few years, the random forest algorithm has been widely used in astronomical classification. There are large differences in the number of different types of variables in astronomy. In oder to solve the problem of population imbalance, we can set \texttt{calss\_weight} = ‘\texttt{balanced\_subsample}’ so that the weight of each class is inversely proportional to the frequency of that class. The random forest algorithm introduces random sampling to avoid overfitting. 
		
		Based on the comparison in Section~\ref{sec:packages3.4}, the 4 parameter values (\emph{logP}, \emph{R$_{21}$}, \emph{R$_{31}$}, \emph{R$_{41}$}) related to the periods of periodic variables in ASAS-SN and OGLE catalogs are reliable. Because the remaining 13 parameter values are either from our calculations or cross-matched from other databases, they are also reliable. Therefore, we directly use 17 features of 241,154 periodic variables (as described in Section~\ref{sec:packages2}) to train the random forest classifier, and classify light curves into 7 superclasses and 17 subclasses. The random forest algorithm can estimate the feature importance based on the mean decrease impurity algorithm (Gini importance) implemented in the \texttt{SCIKIT-LEARN} version of the random forest classifier \citep{Pedregosa2012}. The descriptions and importance of these features is shown in Table~\ref{tab:table3}, the two most important of which is the \emph{logP} and \emph{R$_{21}$}.
		
\begin{figure*}
			\centering
			\includegraphics[width=\textwidth]{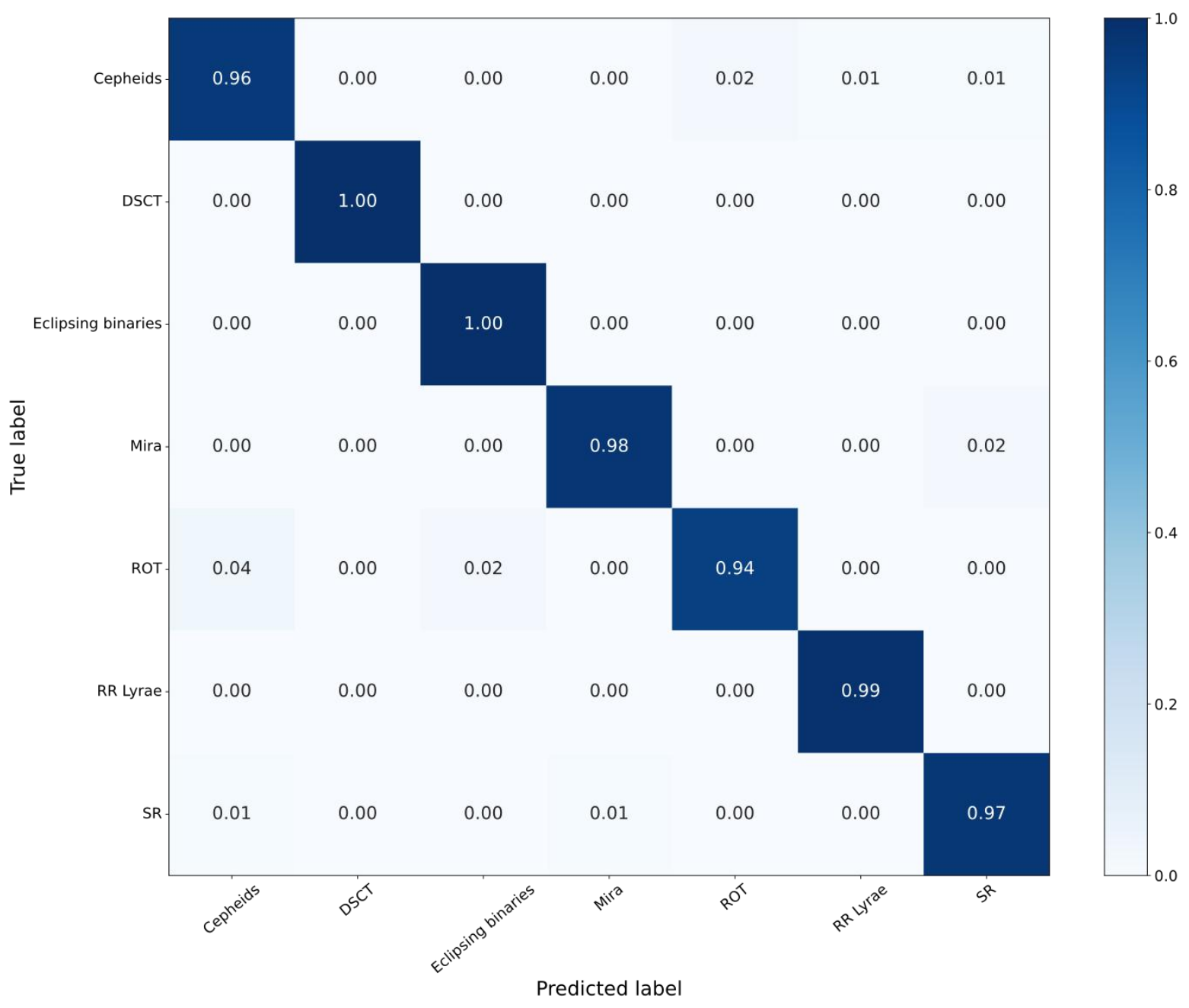}
			\caption{The confusion matrix of superclasses. The y-axis corresponds to the “input” classification given to a variable, while the x-axis represents the “output” prediction obtained from the trained random forest classifier. We show the number if it larger than or equal to 0.01.}
			\label{fig:figure7}
		\end{figure*}

		To evaluate the performance of the classifier, we randomly divide the 241,154 variable star samples into the training set and the testing set. The training set contains 80\% of the population in the original samples. We adopt precision, recall, and F$_{1}$ score to evaluate the performance of the classifier, i.e., 
		\begin{equation}
			precision = \frac{True\,Positive}{True\,Positive + False\,Positive},
		\end{equation}
		
		\begin{equation}
			recall = \frac{True\,Positive}{True\,Positive + False\,Negative},
		\end{equation}
		
		\begin{equation}
			F_{1} = 2 \times \frac{precision \times recall}{precision + recall}.
		\end{equation}
		Table~\ref{tab:table7} shows the precision, recall, and F$_{1}$ scores of 7 superclasses. Fig.~\ref{fig:figure7} shows the confusion matrix of 7 superclasses. The values on the leading diagonal are the recall rate. Each cell value represents the fraction of objects of that true class (rows) assigned to the predicted classes (columns). It is very nice that the recall rate of superclasses ranges from 0.94 to 1.
		
		\begin{table}
			\centering
			\caption{The classification quality of the trained random forest classifier with superclasses.}
				\begin{tabular}{lccc} 
					\hline
					Class & Precision & Recall & F$_{1}$ score\\
					\hline
					Cepheids & 0.902 & 0.963 & 0.931\\
					Delta Scuti & 0.978 & 1.000 & 0.989\\
					Eclipsing binaries & 0.998 & 0.996 & 0.997\\
					Mira & 0.939 & 0.982 & 0.960\\
					ROT & 0.932 & 0.943 & 0.938\\
					RR Lyrae & 0.996 & 0.990 & 0.993\\
					SR & 0.993 & 0.970 & 0.981\\
					\hline
				\end{tabular}
			\label{tab:table7}
		\end{table}
		
\begin{figure*}
			\centering
			\includegraphics[width=\textwidth]{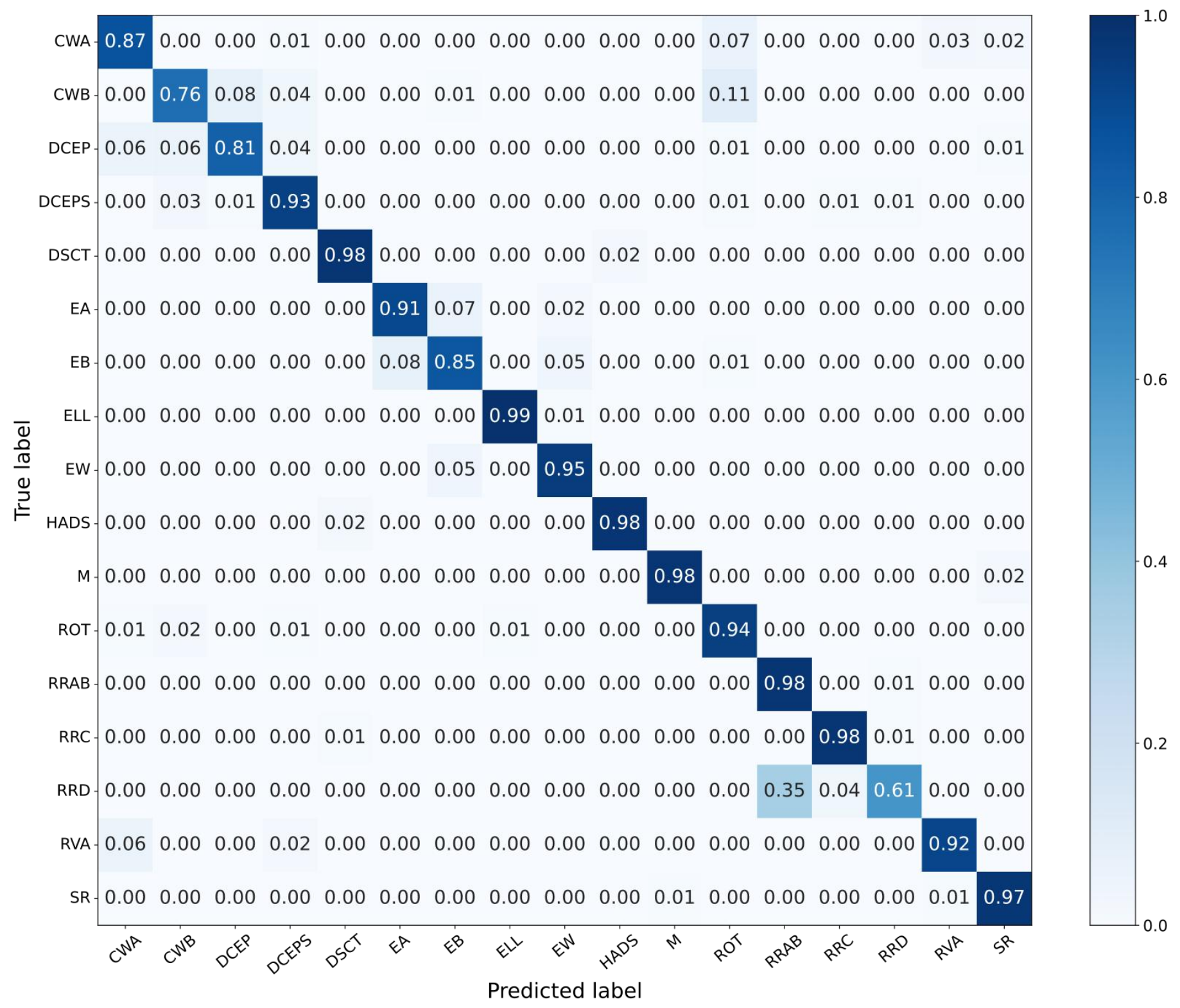}
			\caption{The confusion matrix of subclasses. The y-axis corresponds to the “input” classification given to a variable, while the x-axis represents the “output” prediction obtained from the trained random forest classifier. We show the number if it larger than or equal to 0.01.}
			\label{fig:figure8}
		\end{figure*}

		Table~\ref{tab:table8} shows the precision, recall, and F1 scores of the 17 subclasses. Fig~\ref{fig:figure8} shows the confusion matrix of the 17 subclasses. The recalls of the subclasses are lower than that of superclasses, but most of them are above 90\%. The recall rate of the RRD class is only 0.61, and 35\% of the RRD variables are predicted to be RRAB variables. RRAB, RRC, and RRD Lyrae are three subclasses of RR Lyrae. It is difficult to distinguish the variability characteristics of each subclass under the same superclass. In many works, RR Lyrae are only divided into two subclasses, RRAB and RRC Lyrae. There are only 605 RRD variables in our samples, accounting for 0.25\% of the total variables, which leads to insufficient learning of the classifier and the low recall rate of the RRD class. Due to the rarity of RRD variables across the sky and the inconsistency of classification criteria of different survey projects, it is difficult to collect enough and reliable photometric data of RRD variables.
		
		\begin{table}
			\caption{The classification quality of the trained random forest classifier with subclasses.}
				\centering
				\begin{tabular}{lccc} 
					\hline
					Class & Precision & Recall & F$_{1}$ score\\
					\hline
					CWA	& 0.526	& 0.866	& 0.654\\
					CWB	& 0.510	& 0.757	& 0.609\\
					DCEP & 0.936 & 0.813 & 0.870\\
					DCEPS & 0.867 & 0.928 & 0.896\\
					DSCT & 0.927 & 0.979 & 0.952\\
					EA & 0.950 & 0.908 & 0.929\\
					EB & 0.750 & 0.855 & 0.799\\
					ELL	& 0.969	& 0.985	& 0.977\\
					EW	& 0.970	& 0.946	& 0.958\\
					HADS & 0.978 & 0.979 & 0.978\\
					Mira & 0.939 & 0.982 & 0.960\\
					ROT	& 0.932	& 0.943	& 0.938\\
					RRAB & 0.989 & 0.982 & 0.986\\
					RRC	& 0.988	& 0.979	& 0.984\\
					RRD	& 0.557	& 0.609	& 0.582\\
					RVA	& 0.488	& 0.923	& 0.638\\
					SR & 0.993 & 0.970 & 0.981\\
					\hline
				\end{tabular}
			\label{tab:table8}
		\end{table}
		
		\begin{table}
			\caption{The classification and period comparisons between our own results and ASAS-SN, OGLE catalogs for periodic variable stars (the periods with differences less than $\pm$5\% are considered to be the same).}
			\centering
			\begin{tabular}{l|c|cccc} 
				\hline
				ASAS-SN \& OGLE	& \multicolumn{5}{c}{This work}\\
				\hline
				\multirow{3}{*}{Class} & \multirow{3}{*}{Aperiodic} & \multicolumn{4}{c}{Periodic}\\
				\cline{3-6}
				&  & \multirow{2}{*}{same type} & same type & \multirow{2}{*}{different type} &  different type\\
				&  &  & different period &  & same period\\
				\hline
				Cepheids (2000)	& 0.6\% & 92.0\% & 4.3\% & 8.0\% & 4.8\%\\
				Delta Scuti (2000) & 3.7\% & 93.3\%	& 4.0\% & 6.7\% & 0.0\%\\
				EA (2000) & 17.9\% & 88.7\% & 15.7\% & 11.3\% & 7.9\%\\
				EB (2000) & 2.2\% & 81.7\%	& 0.6\% & 18.3\% & 8.0\%\\
				ELL (2000) & 0.1\% & 72.8\%	& 1.0\% & 27.2\% & 0.3\%\\
				EW (2000) & 0.8\% & 66.9\% & 0.4\% & 33.1\% & 3.6\%\\
				Mira (2000) & 4.2\%	& 98.6\% & 10.7\% & 1.4\% & 1.2\%\\
				ROT (2000) & 8.0\% & 87.6\%	& 9.6\% & 12.4\% & 4.7\%\\
				RRAB (2000)	& 2.4\%	& 97.4\% & 0.2\% & 2.6\% & 1.8\%\\
				RRC (2000)	& 0.7\% & 73.2\% & 0.0\% & 26.8\% & 1.6\%\\
				RRD (605)	& 5.0\% & 62.1\% & 0.2\% & 37.9\% & 30.8\%\\
				SR (2000) & 18.8\% & 86.8\%	& 14.3\% & 13.2\% & 2.5\%\\
				\hline
			\end{tabular}
			\label{tab:table9}
		\end{table}
		
		\begin{table}
			\caption{The classification and period comparisons between our own results and ASAS-SN, OGLE catalogs for periodic RRD variables (the periods with differences less than $\pm$5\% are considered to be the same).}
			\centering
			\begin{tabular}{l|cc|cc} 
				\hline
				\multirow{3}{*}{RRD variables} & \multicolumn{4}{c}{This work}\\
				\cline{2-5}
				& \multicolumn{2}{c}{same type} & \multicolumn{2}{c}{different type}\\
				\cline{2-5}
				& same period & different period & same period & different period\\
				\hline
				ASAS-SN (233)	& 23.2\% & 0.0\% & 74.2\% & 2.6\%\\
				OGLE (342)	& 88.3\% & 0.3\% & 1.2\% & 10.2\%\\
				\hline
			\end{tabular}
			\label{tab:table10}
		\end{table}
		
		\begin{table*}
			\caption{The percentage of meeting 6 criteria for irregular variable stars from ASAS-SN.}
				\centering
				\begin{tabular}{lccccc} 
					\hline
					Superclass & Subclass &	Meet \hyperlink{criterion1-1}{criterion 1-1} & Meet \hyperlink{criterion1-2}{criterion 1-2} & Meet \hyperlink{criterion2-1}{criterion 2-1} & Meet \hyperlink{criterion2-2}{criterion 2-2}\\
					\hline
					\multirow{2}{*}{Irregular variables} & L (48964) &	0.5\%	& 0.1\%	& 0.4\% &	99.0\%\\
					& YSO (1364) & 10.4\% & 0.4\% & 0.5\% & 88.6\%\\
					\hline
				\end{tabular}
			\label{tab:table11}
		\end{table*}
		
		\begin{table}
			\caption{The classification comparisons between our own results and ASAS-SN catalog for irregular variable stars.}
				\centering
				\begin{tabular}{l|c|cccc} 
					\hline
					ASAS-SN & \multicolumn{5}{c}{This work}\\
					\hline
					\multirow{2}{*}{Class} & \multirow{2}{*}{Aperiodic} & \multicolumn{4}{c}{Periodic}\\
					\cline{3-6}
					&  & SR & Cepheids & ELL & ROT\\
					\hline
					L (2000) & 51.4\% & 87.6\% & 8.5\% & 3.4\% & ---\\ 
					YSO (1364) & 38.0\% & 53.4\% & 13.3\% & 5.1\% & 23.0\% \\
					\hline
				\end{tabular}
			\label{tab:table12}
		\end{table}
		
		\subsection{Classification accuracy comparison with ASAS-SN $\&$ OGLE}
		\label{sec:packages4.2}
		After evaluating the performance of the classifier, we train a random forest model using the features of all 241,154 variable star samples and obtain the final classifier. In Section~\ref{sec:packages3.4}, we derive the periods for 2000 varibales (605 samples for RRD variables) of different classes. To test the reliability of the classifier, we calculate the 4 features (\emph{logP}, \emph{R$_{21}$}, \emph{R$_{31}$}, \emph{R$_{41}$}) for these varibales to replace the initial values from ASAS-SN and OGLE catalogs and other features remain unchanged (as described in Section~\ref{sec:packages2}). Ultimately, we use the final classifier to predict variability types. Assuming that the results of the ASAS-SN and OGLE catalogs are 100\% correct and the periods with differences less than $\pm$5\% are regarded the same, we compare our results with ASAS-SN and OGLE catalogs. Table~\ref{tab:table9} shows the classification accuracy and period accuracy. Only small numbers of periodic variable stars are considered to be aperiodic by our method. The classification accuracy are generally above 82\%. For $\sim$ 27\%-33\% ELL, EW, and RRC variables, because our 6 criteria cannot distinguish them very well, the periods are mis-derived, which further leads to mis-classifications. For $\sim$ 10\%-16\% of EA, Mira, ROT and SR variables, we predict the same types as ASAS-SN but the periods are different, which indicate that other features of these variables such as amplitude and \emph{G$_{BP}$ - G$_{RP}$} may be more important than period. For about 8\% of EA and EB variables, we derive the same periods as ASAS-SN but the prediction types are different, which indicate that the variablity types of classifying eclipsing binaries according to the shapes of the light curves may be ambiguous. For RRD variables, we find an interesting case. The period accuracy can reach 92.7\%, but the classification accuracy is only 62\%. We examine the prediction types of those 30.8\% RRD variables and find that they are almost entirely predicted as RRAB variables. Moreover, We use the classifier to classify the RRD variables (605 $\times$ 95\% $\simeq$ 575) from ASAS-SN and OGLE respectively, the results are shown in Table~\ref{tab:table10}. We find that the prediction types of 88.6\% RRD variables from OGLE are correct. However, only 23.2\% RRD variables in ASAS-SN are correctly predicted. \cite{Kim2021} also obtained similar results. The figure 10 in their paper showed that 91\% RRD variables from ASAS-SN were classified as RRAB variables by the model which was trained based on light curves from OGLE \emph{I}-band and EROS-2 \emph{B$_{E}$}-band. The low classification accuracy of RRD variables may be caused by insufficient learning of the classifier, or different classification criteria between ASAS-SN and OGLE. The values in Table~\ref{tab:table9} shows that our period determination method is completely feasible, and the performance of our final classifier is acceptable.

		\subsection{Comparison with irregular variables from  ASAS-SN}
		\label{sec:packages4.3}
		
\begin{figure*}
			\centering
			\includegraphics[width=\textwidth]{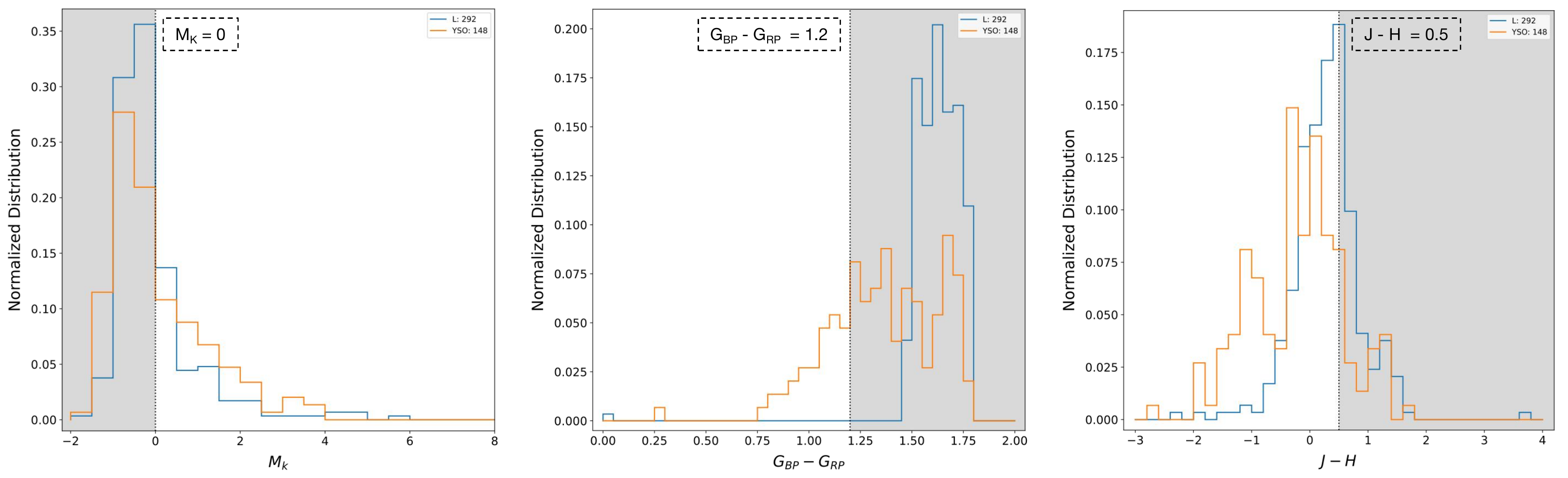}
			\caption{The distributions of L and YSO variables in \emph{M$_{k}$}, \emph{G$_{BP}$} - \emph{G$_{RP}$} and \emph{J - H} diagrams. The numbers of L and YSO variables meeting \protect\hyperlink{criterion1}{criterion 1} are shown in the legend. The region meeting \protect\hyperlink{criterion1-2}{criterion 1-2} is shaded in gray.}
			\label{fig:figure9}
		\end{figure*}

\begin{figure*}
			\centering
			\includegraphics[width=\textwidth]{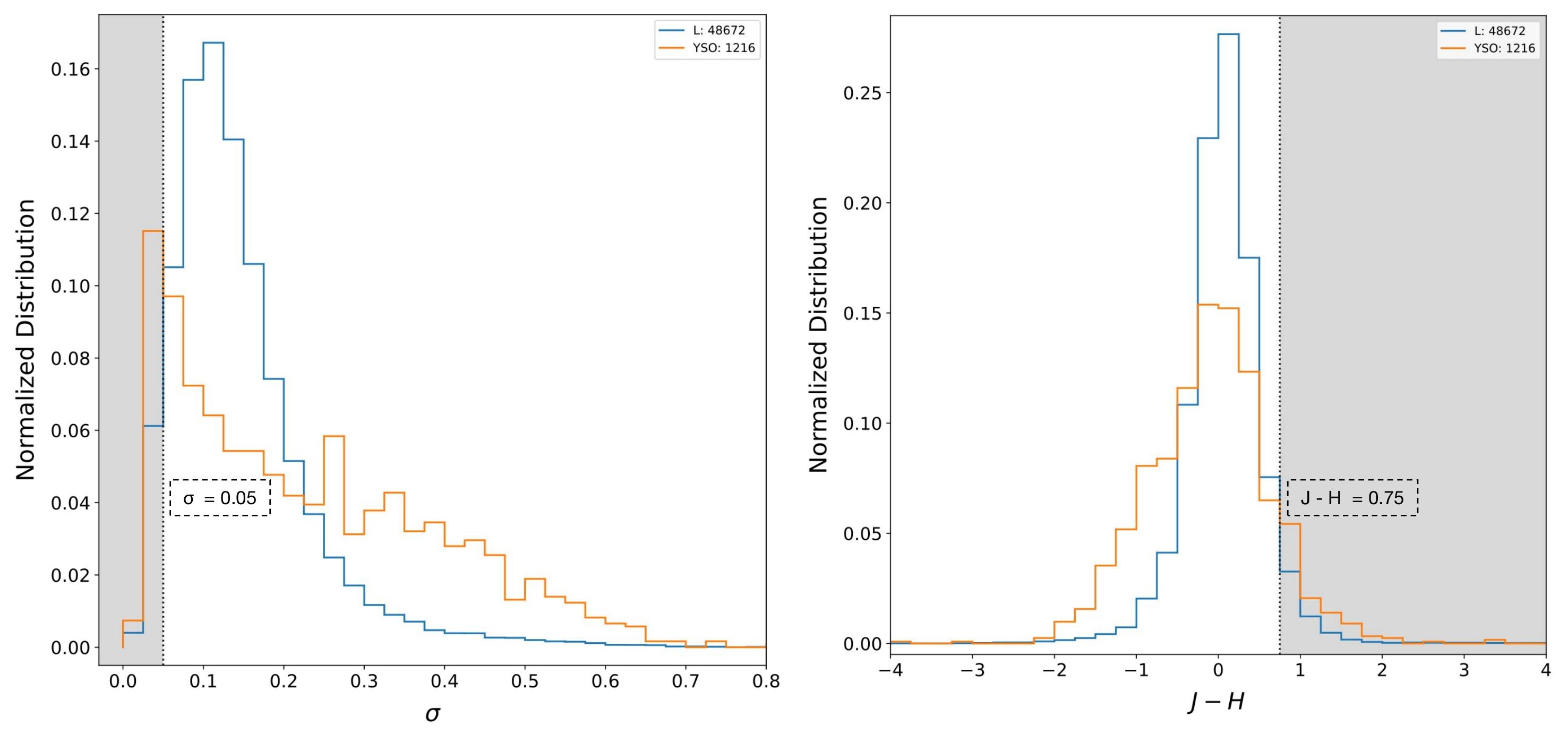}
			\caption{The distributions of L and YSO variables in \emph{$\sigma$} and \emph{J - H} diagrams. The numbers of L and YSO variables meeting \protect\hyperlink{criterion2}{criterion 2} are shown in the legend. The region meeting \protect\hyperlink{criterion2-1}{criterion 2-1} is shaded in gray.}
			\label{fig:figure10}
		\end{figure*}

		In addition to periodic variable stars, we also select 50,328 irregular variables with \emph{classification probability} $\ge$ 0.6 (96.5\% samples with \emph{classification probability} $\ge$ 0.9) and \emph{lksl\_statistic} $\leq$ 0.5 from the ASAS-SN Variable Star Database. For all irregular variables, we calculate the \emph{ALH}, \emph{M$_{k}$} and $\sigma$ and obtain the \emph{G$_{BP}$} - \emph{G$_{RP}$}, \emph{J} - \emph{H} from ASAS-SN catalog, the 5 parameters help to test the efficiency of 6 criteria in our period determination method. The parameter distributions and the ratios of L-type and YSO-type variables meeting the 6 criteria are shown in Fig.~\ref{fig:figure9},~\ref{fig:figure10} and Table~\ref{tab:table11}, without considering the comparisons between the aliased periods and the possible periods and between FAP and FAP1 or FAP2. Although the distributions of L-type and YSO-type variables are not significantly different in Fig.~\ref{fig:figure9} and \ref{fig:figure10}, $\sim$ 99\% of L-type variables and $\sim$ 89\% of YSO-type variables meet \hyperlink{criterion1-2}{criterion 1-2} and \hyperlink{criterion2-2}{2-2}. Therefore, we don't need to worry about irregular variables mixing into eclipsing binaries by setting a lower FAP1 corresponding to 1 $\times$ 10$^{-2}$ for eclipsing binaries.
		
		To test the efficiency of our method and classifier, we first calculate the \emph{MAD}, \emph{IQR}, \emph{A} and obtain the \emph{W1} $-$ \emph{W2}, \emph{W$_{RP}$}, \emph{W$_{JK}$} from ASAS-SN catalog for 2000 L-type variables and 1364 LSO-type vartibales. After using the our method to derive the periods, the 4 features (\emph{logP}, \emph{R$_{21}$}, \emph{R$_{31}$}, \emph{R$_{41}$}) related to the period are calculated. We ultimately use the final classifier to predict variability types.
		
		Table~\ref{tab:table12} only shows the predicted variability types because the vast majority of irregular variables in the ASAS-SN catalog have no period. 51\% of L-type variables are considered aperiodic by the new method, and 88\% of the L-type variables that are regarded as periodic are finally classified as SR variables. 38\% of YSO-type variables are considered as aperiodic, and 53\% of the YSO-type variables that are regarded as periodic are finally predicted to be SR variables. This indicates that the periodicity of L-type and YSO-type variables is indeed not obvious. However, 9\% of L-type variables and 13\% of YSO-type variables are predicted to be Cepheids, 3\% of L-type variables and 5\% of YSO-type variables are predicted to be ELLs, and 23\% of YSO-type variables are predicted to be ROTs. Since 99\% of L-type variables and 89\% of YSO-type variables meet \hyperlink{criterion1-2}{criteria 1-2} and \hyperlink{criterion2-2}{2-2}, for 970 (2000 $\times$ 0.99 $\times$ 0.49 $\simeq$ 970) L-type variables and 753 (1364 $\times$ 0.89 $\times$ 0.62 $\simeq$ 753) YSO-type variables, the corresponding frequency peak values of the best matched periods given by the GLS algorithm exceed FAP2 (10$^{-15}$) and are not eliminated by 10 aliased periods. Therefore, we can believe that these 87 (2000 $\times$ 0.99 $\times$ 0.49 $\times$ 0.09 $\simeq$ 87), 29 (2000 $\times$ 0.99 $\times$ 0.49 $\times$ 0.03 $\simeq$ 29), 98 (1364 $\times$ 0.89 $\times$ 0.62 $\times$ 0.13 $\simeq$ 98),  38 (1364 $\times$ 0.89 $\times$ 0.62 $\times$ 0.05 $\simeq$ 38), 173 (1364 $\times$ 0.89 $\times$ 0.62 $\times$ 0.23 $\simeq$ 173) samples are periodic variables to a great extent.
		
		\section{Classification of Variables in ZTF DR2}\label{sec:packages5}
		
		Our period determination method and random forest classifier are based on the light curves from the ASAS-SN and OGLE databases. To further test the reliability of our method and classifier, we utilize the new period determination method and the final classifier on the variables from ZTF DR2, and compare the results with those in \cite{Chen2020}. 
		
		\cite{Chen2020} determined periods for 781,602 variable stars of ZTF DR2, ranging from 0.025 d to 1000 d, including 745,896 variables in the \emph{g}-band and 765,958 variables in the \emph{r}-band. They classified these variables into 11 subclasses (CEP I, CEP II, DSCT, Mira, RRab, RRc, SR, EA, EW, BY Dra, RS CVn) based on the specific distributions of different variables in multi-dimensional space. They found 621,702 new variables, 74\% of all variables were located at Galactic latitudes $|$b$|$ $<$ 10$^{\circ}$, and the coverage of the north warp and the disk's  edge at distances of 15-20 kpc was significantly better than previously, which contribute to studying the physics of variables and the disk structure and evolution of the Milky Way. The periodic variables catalog and light curves of ZTF DR2 can be accessed from Chen’s homepage\footnote{\url{http://variables.cn:88/ztf/}} and data sharing platform “Zenodo”\footnote{\url{https://zenodo.org/record/3886372}}. 
		
		We directly use the photometric data of 781,602 variables sorted out by \cite{Chen2020}. First, we extract and cross-match 13 featrues ($\sigma$, \emph{ALH}, \emph{M$_{s}$}, \emph{M$_{k}$}, \emph{MAD}, \emph{IQR}, \emph{A}, \emph{G$_{BP}$} - \emph{G$_{RP}$}, \emph{J} - \emph{K$_{s}$}, \emph{J} - \emph{H}, \emph{W1} $-$ \emph{W2}, \emph{W$_{RP}$}, \emph{W$_{JK}$}) for each variables of two bands. Then, we derive periods for variables with 13 features in two bands using the methods described in Sections~\ref{sec:packages3.2} and~\ref{sec:packages3.3}, the 4 features (\emph{logP}, \emph{R$_{21}$}, \emph{R$_{31}$}, \emph{R$_{41}$}) related to the period are calculated. Next, we classify them using the final classifier. The number of variables with all features in the \emph{g}-band is 279,670 and that in the \emph{r}-band is 410,726. The final results preferentially use the periods and variability types in the \emph{r}-band, which has a total of 437,080 variables. 
		
		Due to different name schemes, we adopt the variability type matching scheme in Table~\ref{tab:table13} to compare our periods and variability types with the results of \cite{Chen2020}, and the periods with differences less than $\pm$5\% are regarded as the same. Table~\ref{tab:table14} shows the comparison results. There is a high proportion of being aperiodic or missing features for these variability types (DSCT, RRab, RRc, SR, EA, EW, BY Dra, RS CVn). First, the number of observation points per light curve in ZTF DR2 is less than that in the ASAS-SN Variable Star Database, and  the proportion of variables ragarded as aperiodic by us will be higher. Second, the FAP2 (10$^{-15}$) has stricter periodicity constraints than the FAP (10$^{-3}$) set by \cite{Chen2020}, and the sources without obvious periodicity are excluded. Third, we only analyse the variables with all features, and the sources lacking any one of the features are excluded. 
		
		Except for EW and SR variables, the classification accuracy and period accuracy of other types of variables are above $\sim$ 65\%, and the two comparision results can even reach above 80\% for CEP I, DSCT, EA, and RRAB variables. The values in Table~\ref{tab:table14} are generally lower than the values in Tables~\ref{tab:table5} and~\ref{tab:table9} for various reasons. First, the ASAS-SN and OGLE databases generally have more observation points per light curve and higher classification accuracy. Second, our method cannot select ELL, EW, and RRC variables and derive periods very well (seeing Tables~\ref{tab:table4} and~\ref{tab:table5}). Third, the values in Tables~\ref{tab:table5} and ~\ref{tab:table9} are calculated on the assumption that the ASAS-SN catalog and the results of \cite{Chen2020} about ZTF DR2 are 100\% correct. Finally, the band difference between ASAS-SN and ZTF may be a factor for this result. Our method and classifier are mainly developed with ASAS-SN \emph{V}-band light curves. The light curves of variables from ZTF DR2 are mainly in the \emph{r}-band. However, there is not study to prove the existence of band effects between ASAS-SN and ZTF. The classification consistency based on different bands need to be explored. Similarly, it could be problematic to combine data from ASAS-SN and OGLE since they observe in different bands, but such an effect does not constitute a major influence according to our results.
		
		\begin{table}
			\centering
			\caption{Variability types matching scheme between  \protect\cite{Chen2020} and ASAS-SN catalog.}
				\begin{tabular}{cc} 
					\hline
					\multirow{2}{*}{ZTF DR2 Class} & \multirow{2}{*}{ASAS-SN Class}\\
					& \\
					\hline
					\multirow{2}{*}{BY Dra \& RS CVn} & ROT\\
					& ELL\\
					\hline
					\multirow{2}{*}{CEP I} & DCEP\\
					&  DCEPS\\
					\hline
					\multirow{3}{*}{CEP II} & CWA\\
					& CWB\\
					& RVA\\
					\hline
					\multirow{2}{*}{DSCT} & HADS\\
					& DSCT\\
					\hline
					\multirow{2}{*}{EA} & EA\\
					& EB\\
					\hline
					\multirow{2}{*}{EW} & EB\\
					& EW\\
					\hline
					Mira & Mira\\
					\hline
					RRab & RRAB\\
					\hline
					\multirow{2}{*}{RRc} & RRC\\
					& RRD\\
					\hline
					SR & SR\\
					\hline
				\end{tabular}
			\label{tab:table13}
		\end{table}
		
		\begin{table*}
			\centering
			\caption{The classification and period comparisons between our results and those of  \protect\cite{Chen2020} for periodic variable stars (the periods with differences less than $\pm$5\% are considered to be the same).}
				\begin{tabular}{c|c|cccc} 
					\hline
					ZTF DR2 & \multicolumn{5}{c}{This work}\\
					\hline
					\multirow{3}{*}{Class} & \multirow{2}{*}{Aperiodic or}  & \multicolumn{4}{c}{Periodic}\\
					\cline{3-6}
					& \multirow{2}{*}{Feature omission} & \multirow{2}{*}{same prediction} & \multirow{2}{*}{same period} & same period & \multirow{2}{*}{different prediction} \\
					&  &  &  & same predictions & \\
					\hline
					BY Dra (84697) & 40.3\% & 88.1\% & 64.4\% & 60.4\% & 11.9\%\\
					\multirow{2}{*}{RS Cvn (81393)} &  \multirow{2}{*}{36.9\%} &  \multirow{2}{*}{62.0\%} &  \multirow{2}{*}{70.3\%} &  \multirow{2}{*}{46.1\%} & 38.0\% (RRC, EB, DSCT, EW,\\
					& & & & &  CWB, HADS, CWA)\\
					\hline
					CEP I (1262) & 12.4\% & 83.8\% & 98.2\% & 83.6\% & 16.2\%\\
					CEP II (358) & 22.6\% & 69.7\% & 81.6\%	& 63.9\% & 30.3\% (DCEP, EB, DCEPS)\\
					\hline
					DSCT (16709) & 47.6\% & 95.4\% & 84.0\%	& 83.9\% & 4.5\%\\
					\hline
					EA (49943) & 45.7\% & 86.6\% & 79.3\% & 76.8\% & 13.4\%\\
					EW (369707) & 48.0\% & 48.2\% & 49.6\% & 47.1\% & 51.8\% (RRC, HADS, DSCT)\\
					\hline
					Mira (11879) & 17.1\% &	77.9\% & 68.6\%	& 55.0\% & 22.2\%\\
					\hline
					RRab (32518) & 69.1\% & 80.4\% & 87.1\%	& 79.7\% & 19.6\%\\
					RRc (13875)	& 49.5\% & 63.1\% & 65.4\% & 62.5\% & 37.0\% (EW)\\
					\hline
					SR (119261)	& 33.8\% & 54.2\% & 53.4\% & 41.7\% & 45.9\% (DCEP, DCEPS, ELL, ROT)\\
					\hline
				\end{tabular}
			\label{tab:table14}
		\end{table*}

		\section{Conclusions}
		\label{sec:packages6}
		
		We develop a new method to determine the periods of variables. By combining the statistical parameters of the light curves, the colors of the variables, the variables are initially divided into eclipsing binaries and NEB variables. By using the window function, the GLS algorithm, and two FAP levels, the periods of eclipsing binaries and NEB variables are obtained through different processing steps. In comparison with the ASAS-SN catalog, the period accuracy of different types of variables is 70\%-99\% (seeing Table~\ref{tab:table5}). In comparison with the results of \cite{Chen2020} for ZTF DR2, the period accuracy of the EW and SR variables is $\sim$ 50\% and 53\%, respectively. The period accuracy of other types of variable stars is 65\%-98\%. This method can return the true periods for most variable stars. However, for the EW, ELL, RRC variables or similar types of variables (BY Dra, RS CVn), the method does not achieve  very high period accuracy due to the highly overlapping parameter distributions of these types of variables.
		
		We select 241,154 periodic variables with high classification accuracy from the ASAS-SN and OGLE databases. We extract 11 features from each light curve and cross-match 6 features from the Gaia Early DR3, ALLWISE, and 2MASS catalogs to train the random forest classifier and classify these variables into 7 superclasses and 17 subclasses. By testing the performance of the classifier, 35\% of the RRC variables are predicted as RRAB variables,76\%-99\% of other types of variables are predicted as the true types. Compared with the ASAS-SN and OGLE catalogs and the results of \cite{Chen2020} for ZTF DR2, the classification accuracy is generally above 82\% and 70\%, respectively.
		
		At present, there are many classification models with their own characteristics. These models cannot be applied directly on the data from the CNEOST or the WFST. It is necessary to develope a new period determination method and establish a new random forest classifier to efficiently process the large data sets. In the future, we plan to use the period determination method and the classifier to identify and classify variable stars in the CNEOST and WFST databases. We will continuously improve the period determination method and classifier performance to accurately identify variable stars and discover interesting sources.
		
		\section*{acknowledgments}
		This work is supported by the Strategic Priority Research Program of Chinese Academy of Sciences (No. XDB 41000000), China Postdoctoral Science Foundation (2021M703099), and the Fundamental Research Funds for the Central Universities. The authors gratefully acknowledge the support of Cyrus Chun Ying Tang Foundations. Special thanks to Prof. T. Jayasinghe for answering our questions via email. We thank the team of Prof. Chen for directly using the photometric data of the ZTF DR2 sorted out  by \cite{Chen2020}.
		
		This work has made use of data from the ASAS-SN Variable Stars Database (\url{https://asas-sn.osu.edu/variables}). ASAS-SN is funded in part by the Gordon and Betty Moore Foundation through grants GBMF5490 and GBMF10501 to the Ohio State University, and also funded in part by the Alfred P. Sloan Foundation grant G-2021-14192. TJ, KZS and CSK are supported by NSF grants AST-1814440 and AST-1908570. B.J.S. is supported by NASA grant 80NSSC19K1717 and NSF grants AST-1920392 and AST-1911074. Support for T.W.-S.H. was provided by NASA through the NASA Hubble Fellowship grant HST-HF2-51458.001-A awarded by the Space Telescope Science Institute (STScI), which is operated by the Association of Universities for Research in Astronomy, Inc., for NASA, under contract NAS5-26555. S.D. acknowledges Project 12133005 supported by National Natural Science Foundation of China (NSFC). 
		
		Development of ASAS-SN has been supported by NSF grant AST-0908816, the Mt. Cuba Astronomical Foundation, the Center for Cosmology and AstroParticle Physics at the Ohio State University, the Chinese Academy of Sciences South America Center for Astronomy (CAS-SACA), the Villum Foundation, and George Skestos. TAT is supported in part by Scialog Scholar grant 24216 from the Research Corporation. Support for JLP is provided in part by FONDECYT through the grant 1151445 and by the Ministry of Economy, Development, and Tourism’s Millennium Science Initiative through grant IC120009, awarded to The Millennium Institute of Astrophysics, MAS.
		
		This work has made use of data from the OGLE Collection of Variable Stars (\url{https://ogledb.astrouw.edu.pl}). The OGLE project has received funding from the European Research Council under the European Community’s Seventh Framework Programme (FP7/2007-2013) / ERC grant agreement no. 246678 to AU. RP is supported by the Foundation for Polish Science through the Start Program.
		
		This publication is based on observations obtained with the Samuel Oschin 48-inch Telescope at the Palomar Observatory as part of the Zwicky Transient Facility project. ZTF is supported by the National Science Foundation under grant AST-1440341 and a collaboration including Caltech, IPAC, the Weizmann Institute for Science, the Oskar Klein Center at Stockholm University, the University of Maryland, the University of Washington, Deutsches Elektronen-Synchrotron and Humboldt University, Los Alamos National Laboratories, the TANGO Consortium of Taiwan, the University of Wisconsin at Milwaukee, and Lawrence Berkeley National Laboratories. Operations are conducted by COO, IPAC, and UW.
		
		This work has made use of data from the European Space Agency (ESA) mission Gaia (\url{https://www.cosmos.esa.int/gaia}), processed by the Gaia Data Processing and Analysis Consortium (DPAC, \url{https://www.cosmos.esa.int/web/gaia/dpac/consortium}). Funding for the DPAC has been provided by national institutions, in particular the institutions participating in the Gaia Multilateral Agreement.
		
		This publication makes use of data products from the Widefield Infrared Survey Explorer, which is a joint project of the University of California, Los Angeles, and the Jet Propulsion Laboratory/California Institute of Technology, funded by the National Aeronautics and Space Administration. 
		
		This publication makes use of data products from the Two Micron All Sky Survey, which is a joint project of the University of Massachusetts and the Infrared Processing and Analysis Center/California Institute of Technology, funded by the National Aeronautics and Space Administration and the National Science Foundation.
		
		~\\
		\emph{Software}: Scikit-learn \citep{Pedregosa2012}, Astropy \citep{AstropyCollaboration2013}, Astrobase \citep[v0.5.3,][]{Bhatti2021}, Gatspy \citep[v0.3,][]{Vanderplas2016}.


		\bibliography{PASPsample631}{}
		\bibliographystyle{aasjournal}
		
		
		
	\end{document}